\documentclass[10pt,conference]{IEEEtran}

\usepackage{amsmath,amssymb,amsfonts, amsthm}
\usepackage{graphicx}
\usepackage{cite}
\usepackage{hyperref}
\usepackage{booktabs}
\usepackage{algorithm}
\usepackage{algpseudocode}
\usepackage{multirow}

\newtheorem{theorem}{Theorem}
\newtheorem{lemma}[theorem]{Lemma}
\newtheorem{corollary}[theorem]{Corollary}
\newtheorem{conjecture}[theorem]{Conjecture}
\theoremstyle{remark}
\newtheorem{remark}[theorem]{Remark}

\usepackage{tikz}
\usetikzlibrary{
    decorations.pathreplacing,  
    arrows.meta,                
    shapes.geometric,           
    calc,                       
    fit,                        
    backgrounds                 
}

\usepackage{quantikz}

\begin{document}

\title{QuIC: A Training-Free Quantum Graph Embedding from Ideal Analysis to Practical Hardware Evaluation}    \author{
\IEEEauthorblockN{Luke Miller, Yugyung Lee}
\IEEEauthorblockA{University of Missouri--Kansas City \\
ljmbm5@umsystem.edu}
}

\maketitle
    
\begin{abstract}
We introduce QuIC, a training-free quantum graph embedding that maps graphs to sorted output distributions via a fixed parameterized circuit. In the ideal one-repetition setting, we prove that the resulting sorted distribution is permutation-invariant and injective on labeled graphs under an irrational-angle condition, yielding completeness on isomorphism classes for the ideal one-repetition exact-arithmetic embedding.

We then use those ideal structural properties to motivate a practical embedding pipeline and study how much of that behavior survives under finite-shot estimation, truncation, realistic noise, transpilation, and hardware execution. The sorted distribution concentrates discriminative signal in a compact head, making fixed-length head truncation an effective practical operating point in the tested regimes. Under noise-model simulation, all tested graph pairs satisfied the study's operational separation criterion, including strongly regular graph pairs that are standard 2-WL stress tests and CFI families used as hard instances for fixed-k WL methods.

A hardware study comprising 14,800 transpiled circuits across 37 CFI families on IBM Heron (\texttt{ibm\_fez}, 156 qubits), including paired one- and two-repetition evaluations, reports empirical separation up to 66 qubits for the tested families under the reported execution protocol, identifies a device-dependent depth limit near 210--250 layers, and characterizes the current practical boundary of the method under the reported execution protocol.
\end{abstract}

\textbf{Note}. This preprint contains the complete manuscript with all mathematical proofs and experimental details in the appendices. A condensed version fitting the IEEE QCE 2026 page limit has been submitted for review.

\section{Introduction}\label{sec:intro}

Graph representations are central to combinatorial optimization,
network analysis, chemistry, and machine learning. A persistent
challenge is constructing representations that are invariant to
vertex relabeling while remaining expressive enough to separate
non-isomorphic structures. Classical expressiveness is commonly
measured against the Weisfeiler--Leman (WL) hierarchy, which bounds
standard message-passing architectures and motivates evaluation on
canonical hard instances such as Cai--F\"{u}rer--Immerman (CFI)
graph pairs. Higher-order methods can exceed 1-WL, but at rapidly
growing computational cost.

Quantum circuits offer an alternative encoding mechanism, but
existing quantum graph methods are limited: some remain
simulation-based, others require training and labeled data, and
still others lack formal guarantees about injectivity or
permutation invariance. A gap persists between mathematically
analyzable quantum graph representations and methods validated on
NISQ hardware.

This paper presents QuIC, a training-free quantum graph embedding
designed to connect ideal analysis with practical execution within
a single framework. QuIC assigns one qubit per vertex, encodes
degree via $R_X$ rotations, encodes edges via $R_{ZZ}$ entangling
gates, and applies global mixer layers before measurement.
Sorting the output distribution removes label dependence, yielding
a graph-level embedding. In the ideal one-repetition setting, we
prove that the resulting sorted distribution is permutation-invariant
and injective on labeled graphs under an irrational-angle condition,
yielding completeness on isomorphism classes for the ideal one-repetition exact-arithmetic embedding.
These ideal structural properties motivate a practical embedding
pipeline, which we then study under the constraints that govern
real execution: finite-shot estimation, fixed-length head truncation,
noise, transpilation, and hardware depth limits.

Unlike WL-style methods, the entangling layer acts in a single
commuting step, encoding the full graph topology globally rather
than through iterative neighborhood refinement. The paper therefore
proceeds in two stages: first, ideal analysis of the one-repetition
circuit; second, empirical characterization of practical realizations
that approximate that ideal object under current hardware constraints.

\subsection{Contributions}\label{sec:intro:contributions}

\textbf{Ideal-theory contributions.}
\begin{enumerate}
    \item A training-free quantum graph embedding with fixed
    parameters and no labeled data requirements.
    \item Proof that the ideal one-repetition sorted output
    distribution is permutation-invariant and injective on labeled
    graphs under an irrational-angle condition, yielding completeness
    on isomorphism classes in that exact regime.
    \item Characterization of the circuit's global encoding structure:
    the entangling layer encodes full-graph topology in a single step,
    and the resulting practical bottleneck is sampling rather than
    information reach.
\end{enumerate}

\textbf{Practical characterization contributions.}
\begin{enumerate}
\setcounter{enumi}{3}
    \item Empirical characterization of the distributional head,
    supporting fixed-length head truncation as a practical operating
    point in the tested shot budgets and graph families.
    \item Noise-model evidence that all tested graph pairs satisfied
    the study's operational separation criterion across exhaustive,
    random, structured, and CFI test suites, including strongly
    regular graphs indistinguishable by 2-WL and CFI pairs
    indistinguishable by $k$-WL for any fixed $k$.
\end{enumerate}

\section{Related Work}\label{sec:related}

Quantum representations of graph-structured data fall into two categories: graph embeddings studied as mathematical objects, and quantum models that process graphs within a learning pipeline.

\subsection{Graph Representations and Expressiveness}
\label{sec:related:expressiveness}

The dominant classical framework for measuring graph representation expressiveness is the Weisfeiler--Leman (WL) hierarchy~\cite{cai1992optimal}. WL-based graph kernels~\cite{shervashidze2011weisfeiler} and message-passing graph neural networks (MPNNs)~\cite{morris2019weisfeiler,xu2018how} are bounded by the 1-WL test. Higher-order $k$-dimensional GNNs exceed this at $O(n^k)$ cost, rendering $k>3$ impractical~\cite{morris2019weisfeiler}. Subgraph counting~\cite{bouritsas2022improving}, topological message passing~\cite{bodnar2021weisfeiler}, and subgraph aggregation~\cite{frasca2022understanding} also surpass 1-WL but at substantial additional cost; recent work proposes finer-grained quantitative expressiveness measures~\cite{zhang2024beyond}. The canonical hard instances for any fixed-$k$ WL method are Cai--F\"{u}rer--Immerman (CFI) pairs, constructed to defeat $k$-variable counting logic~\cite{cai1992optimal}.

Practical isomorphism tools --- nauty/Traces~\cite{mckay2014practical}, bliss~\cite{junttila2007engineering}, and Babai's quasipolynomial algorithm~\cite{babai2016graph} --- solve GI efficiently but return canonical labelings or yes/no decisions, not vectorial embeddings usable for downstream tasks. Complete graph kernels (injective on isomorphism classes) are GI-hard, and none in practical use achieves completeness~\cite{kriege2020survey}.

On the quantum side, graph-state invariants based on Wigner-function measurements have been validated in simulation up to nine nodes and on strongly regular graphs up to 29 nodes, though CFI instances remained out of reach~\cite{Mills2019quantum}. Related work studies efficiently computable invariants under local-complementation equivalence~\cite{burchardt2025foliage}. Both remain simulation-based.

\subsection{Quantum Models for Graph Learning}\label{sec:related:quantum_models}

Quantum kernel methods map classical data into quantum feature space via parameterized circuits, with state inner products defining a kernel~\cite{havlivcek2019supervised,schuld2019quantum}; supervised quantum models have been shown to be kernel methods broadly~\cite{schuld2021quantum}. Neutral-atom processors have been applied to graph algorithms~\cite{dalyac2024graph}, and the GDQC kernel extends this to attributed graphs with expressiveness comparable to GD-WL~\cite{djellabi2025attributed}.

Permutation-invariant quantum machine learning~\cite{mansky2025solving,mansky2025clique} is the most structurally related work: ans\"{a}tze symmetrized under node permutations improve graph classification on tasks including connectivity, bipartiteness, and Hamiltonian path existence. Like QuIC, these exploit permutation invariance as an inductive bias, but they produce \emph{learned} representations optimized on labeled data and subject to barren plateaus~\cite{mcclean2018barren}. QuIC requires no training; its parameters are fixed, the output distribution is the embedding, and completeness is a structural property of the map.

\subsection{Positioning QuIC}\label{sec:related:positioning}

QuIC is not naturally situated within the WL hierarchy~\cite{cai1992optimal,morris2019weisfeiler} and is not an extension in the spirit of beyond-1-WL methods~\cite{bouritsas2022improving,bodnar2021weisfeiler,frasca2022understanding}. Unlike WL-style refinement procedures, QuIC does not operate through iterative neighborhood aggregation; its entangling layer encodes graph structure globally within a single circuit application. For this reason, CFI pairs and other classically difficult instances are used here as empirical stress tests rather than as evidence of a formal separation from WL. Its closest precursors are quantum-walk approaches to graph discrimination, where interacting two-particle walks distinguish all tested non-isomorphic strongly regular graph pairs while non-interacting walks cannot~\cite{gamble2010two,rudinger2013comparing}. QuIC replaces particle dynamics with a parameterized circuit whose entangling layer plays the role of the interaction mechanism.

QuIC is distinct from prior quantum graph methods in both framework and hardware scope. Quantum kernels~\cite{havlivcek2019supervised,schuld2019quantum,schuld2021quantum} and permutation-invariant QML~\cite{mansky2025solving,mansky2025clique} require training, whereas QuIC is a fixed, training-free graph embedding. Prior quantum invariants remain simulation-based~\cite{Mills2019quantum,burchardt2025foliage,djellabi2025attributed}. QuIC's completeness result (Theorem~\ref{thm:injectivity}, Corollary~\ref{cor:complete}) is also unusual: in the ideal one-repetition setting, it yields an invariant and injective embedding on isomorphism classes, an ideal exact-regime guarantee not typically provided by practical graph kernels in common use~\cite{kriege2020survey}. This work then follows that ideal construction through noisy simulation and physical hardware execution.

\section{QuIC Circuit Construction}\label{sec:construction}

QuIC maps a simple undirected graph $G=(V,E)$ with $|V|=k$ vertices to a $k$-qubit circuit, one qubit per vertex. The circuit resembles a QAOA ansatz --- graph-dependent entanglers alternating with mixers --- but parameters are fixed rather than optimized, and the full sorted output distribution serves as the embedding. Three operator families compose the circuit: a degree-encoding layer $U_{\mathrm{enc}}$, an edge-based entangling layer $U_{\mathrm{ent}}$, and a mixing layer $U_{\mathrm{mix}}$, with the entangling and mixing layers repeated for $r$ rounds:
\begin{equation}
    U(G)=\left(U_{\mathrm{mix}}U_{\mathrm{ent}}\right)^r U_{\mathrm{enc}},
    \label{eq:quic_unitary}
\end{equation}
applied to $\lvert 0\rangle^{\otimes k}$. All gate angles are assumed to satisfy $\theta_{\mathrm{enc}}, \theta_{\mathrm{ent}}, \theta_{\mathrm{mix}} \notin \pi\mathbb{Q}$; this condition is formalized in Section~\ref{sec:math}.

\textbf{Encoding layer.} Each qubit receives an $R_X$ rotation proportional to its vertex's normalized degree:
\begin{equation}
    U_{\mathrm{enc}}= \bigotimes_{i=1}^{k} R_X(\phi_i), \qquad \phi_i=
        \theta_{\mathrm{enc}}\dfrac{d_i}{d_{\max}}, \quad d_{\max}>0
    \label{eq:encoding}
\end{equation}
where $d_i=\deg(v_i)$ and $d_{\max}=\max(d)$. Degree information is implicitly present in the entangling layer through the number of $R_{ZZ}$ gates per qubit; the encoding layer makes it explicitly available in the initial amplitudes, ensuring the entangler operates on a state already differentiated by local connectivity. This double encoding is essential: the injectivity proof (Section~\ref{sec:math}) recovers vertex degrees from encoder amplitudes as its first step, and without per-vertex differentiation, regular graphs produce uniform initial states from which edge structure cannot be recovered.

\textbf{Entangling layer.} For each edge $(v_i,v_j)\in E$:
\begin{equation}
    U_{\mathrm{ent}}= \prod_{(i,j)\in E} R_{ZZ}^{(i,j)}(\theta_{\mathrm{ent}}).
    \label{eq:entangler}
\end{equation}
$R_{ZZ}$ is diagonal in the computational basis, contributing a phase proportional to the cut function of $G$ at each bitstring (Section~\ref{sec:math}). Since $R_{ZZ}$ gates on disjoint pairs commute, the layer is order-invariant.

\textbf{Mixing layer.} A uniform $R_X$ rotation on every qubit:
\begin{equation}
    U_{\mathrm{mix}}= \bigotimes_{i=1}^{k} R_X(\theta_{\mathrm{mix}}).
    \label{eq:mixer}
\end{equation}
The mixer redistributes amplitude across computational-basis states, propagating structural information into the measured distribution.

\textbf{Output.} The QuIC embedding is the sorted distribution $\tilde{p}(G)$ obtained by arranging
\begin{equation}
    p_G(x)=\left|\langle x \mid U(G)\mid 0\rangle^{\otimes k}\right|^2,
    \qquad x\in\{0,1\}^k,
    \label{eq:output_distribution}
\end{equation}
in non-increasing order. Sorting removes dependence on vertex labeling; permutation invariance is formalized in Section~\ref{sec:math}.

\textbf{Computational profile.} The logical QuIC circuit uses $|V(G)|$ qubits and $|E(G)|$ two-qubit entangling gates per repetition, so its pre-transpilation serial depth is $O(|E(G)|)$ under a naive schedule and may be reduced by parallel gate packing when graph structure and hardware connectivity permit. Exact statevector simulation remains exponential in $|V(G)|$, as with any full-amplitude simulation of an $n$-qubit circuit. In practical finite-shot settings, however, QuIC is compared through sampled output counts and top-$k$ extraction rather than full-state sorting, so the dominant post-processing cost scales with the number of collected samples or observed bitstrings rather than the full $2^{|V(G)|}$ output space. These costs are not presented as a claim of computational advantage over classical graph isomorphism or graph kernels; rather, they clarify that the main practical bottlenecks in the present study arise from sampling, transpilation variability, and hardware noise rather than from circuit construction itself.

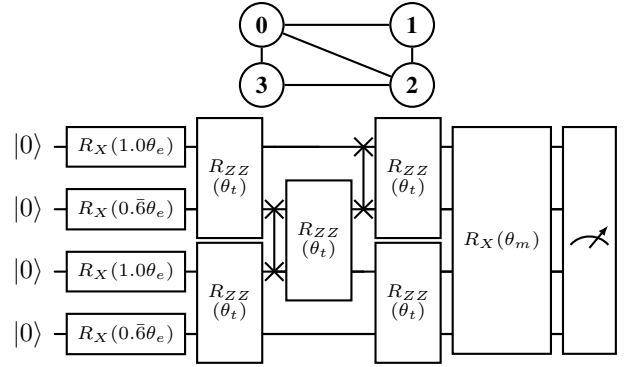
\begin{figure}
    \centering
    \begin{minipage}{1\linewidth}
    \centering
      \begin{tikzpicture}[
        vertex/.style={circle,draw=black,thick,minimum size=0.58cm,
                       inner sep=1pt,font=\small\bfseries}
      ]
        \node[vertex] (n0) at (0,  0.8) {0};
        \node[vertex] (n1) at (2,0.8) {1};
        \node[vertex] (n2) at (2,0)   {2};
        \node[vertex] (n3) at (0,  0)   {3};
        \draw[thick] (n0)--(n1) (n1)--(n2)
                     (n2)--(n3) (n0)--(n3) (n0)--(n2);
      \end{tikzpicture}
    \end{minipage}    
    \begin{minipage}{\linewidth}
        \begin{quantikz}[row sep={0.83cm,between origins}, column sep=0.16cm]
            \lstick{\ket{0}}
                & \gate{\scriptstyle{R_X(1.0\theta_e)}}
                & \gate[2]{\substack{R_{ZZ}\\  (\theta_t)}}
                & \qw
                & \qw
                & \swap{1}
                & \gate[2]{\substack{R_{ZZ}\\  (\theta_t)}}
                & \gate[4]{\scriptstyle{R_X(\theta_m)}}
                & \meter[4]{}
                \\
            \lstick{\ket{0}}
                & \gate{\scriptstyle{R_X(0.\bar{6}\theta_e)}}
                & \qw
                & \swap{1}
                & \gate[2]{\substack{R_{ZZ}\\  (\theta_t)}}
                & \targX{}
                & \qw
                & \qw
                & \qw
                \\
            \lstick{\ket{0}}
                & \gate{\scriptstyle{R_X(1.0\theta_e)}}
                & \gate[2]{\substack{R_{ZZ}\\  (\theta_t)}}
                & \targX{}
                & \qw
                & \qw
                & \gate[2]{\substack{R_{ZZ}\\  (\theta_t)}}
                & \qw
                & \qw
                \\
            \lstick{\ket{0}}
                & \gate{\scriptstyle{R_X(0.\bar{6}\theta_e)}}
                & \qw
                & \qw
                & \qw
                & \qw
                & \qw
                & \qw
                & \qw
                \\
        \end{quantikz}
    \end{minipage}
    \caption{Example four-vertex graph (top) and its one-layer QuIC circuit (bottom). Degree-proportional encoding assigns $0.\bar{6}\theta_e$ to degree-2 vertices and $1.0\theta_e$ to degree-3 vertices. $R_{ZZ}$ gates comprise the entanglement layer. A final, uniform $R_x$ is the mixer. SWAP gates included for figure clarity.}
    \label{fig:construction}
\end{figure}

\textbf{Scope distinction.} We distinguish ideal and practical regimes throughout. The \emph{ideal QuIC embedding} is the full sorted output distribution of the one-repetition circuit under exact arithmetic --- the object for which completeness is proven in Section~\ref{sec:math}. All experiments evaluate \emph{practical QuIC embeddings}, obtained under finite-shot estimation and, on hardware, under transpilation and device noise with vector truncation. These practical studies include both the one-repetition circuit covered by the injectivity theorem and higher-repetition variants, which share the same graph-to-circuit construction but fall outside the current proof once $r>1$.

\subsection{Parameter Selection}\label{sec:construction:grid}

All four parameters ($\theta_{\mathrm{enc}}$, $\theta_{\mathrm{ent}}$, $\theta_{\mathrm{mix}}$, $r$) were selected via systematic sweeps on ER, BA, and SRG families under noiseless simulation, with final validation on CFI pairs (12--28 qubits). Configurations were ranked by $L_1$ z-score using 64 same-circuit subsample comparisons at 1,024 shots (null) and pairwise non-isomorphic comparisons at 4,096 shots (signal).

\textbf{Repetitions.} The injectivity proof (Theorem~\ref{thm:injectivity}) applies at $r=1$; additional repetitions are not currently covered by the formal guarantee. Empirically, however, $r=2$ consistently produced the largest mean $L_1$ separation across all tested simulation families (TV distances of 0.321, 0.251, 0.323 for ER, BA, SRG vs.\ 0.301, 0.240, 0.312 at $r=1$). Three repetitions were comparable but slightly below $r=2$; four degraded to or below $r=1$ for ER and BA, consistent with additional rounds driving the output toward uniformity. We therefore adopt $r=2$ as the preferred practical operating point in noiseless and noise-model simulation --- an engineering choice that improves finite-shot separation, not a claim that the completeness result extends to repeated circuits. The hardware study then reports both $r=2$ and $r=1$: the former as the practically stronger setting in most tested instances, and the latter as the repetition regime covered by the current theory.

\textbf{Encoding angle.} Performance is most sensitive to $\theta_{\mathrm{enc}}$ (Table~\ref{tab:enc_sweep}). Near $\theta_{\mathrm{enc}} \approx 0$ or $\pi$, qubits approach computational-basis states, the entangler contributes primarily global phase, and degree sensitivity is lost. The optimum $\theta_{\mathrm{enc}} = 2.875 \approx 0.91\pi$ sits just below the $\pi$ boundary: high-degree vertices rotate near $\lvert 1\rangle$ while low-degree vertices retain superposition, ensuring the entangler operates on a structurally differentiated state. Symmetric formulas such as $\theta_v = \pi\, d_v/(d_{\max}+1)$ failed on regular graphs, where uniform initial states eliminate the per-vertex differentiation the injectivity proof requires.

\begin{table}[t]
    \centering
    \caption{Encoding-angle sweep on CFI pairs (12--28 qubits, 4,096 shots).}
    \label{tab:enc_sweep}
    \setlength{\tabcolsep}{7pt}
    \renewcommand{\arraystretch}{1.18}
    \begin{tabular}{cccc}
        \hline
        $\theta_{\mathrm{enc}}$ & $\bar{z}$ (mean) & $z_{\min}$ (worst) & Status \\
        \hline
        2.000           & 0.72              & $-$0.77       & Failed            \\
        2.500           & 3.07              &    2.15       & Good              \\
        2.750           & 3.65              &    2.81       & Strong            \\
        \textbf{2.875}  & \textbf{3.79}     & \textbf{2.85} & \textbf{Optimal}  \\
        3.000           & 3.57              &    2.64       & Good              \\
        3.500           & 3.21              & $-$1.67       & Unstable          \\
        4.500           & $-$1.66           & $-$2.51       & Failed            \\
        6.000           & 2.38              &    0.03       & Recovering        \\
        \hline
    \end{tabular}
\end{table}

\textbf{Entangling angle.} Flat across $\theta_{\mathrm{ent}} \in [1.75, 2.25]$ (z-scores vary $<$0.02); we select $\theta_{\mathrm{ent}} = 2.0$.

\textbf{Mixing angle.} $z_{\min}$ rises from 2.01 (no mixer) to 2.44 at $\theta_{\mathrm{mix}} = 0.1$; values above 0.7 destroy separation. Not required but improves worst-case robustness.

\textbf{Selected configuration.} $\theta_{\mathrm{enc}} = 2.875$, $\theta_{\mathrm{ent}} = 2.0$, $\theta_{\mathrm{mix}} = 0.1$, and a preferred simulation operating point of $r = 2$ ($z_{\mathrm{worst}} = 2.85$ on tuning instances). The hardware study additionally reports paired $r=2$ and $r=1$ results to compare the empirically stronger repeated setting against the one-repetition regime covered by the current theory.

\textbf{Sensitivity hierarchy.} Encoding angle is highly sensitive (aliasing near 0 and $\pi$); entangling angle is robust; mixer is beneficial small, destructive large; repetitions peak at $r=2$, degrade at $r=4$. Degree-proportional encoding drives expressiveness; entanglement and mixing amplify it, with diminishing returns at excessive depth.

\section{Mathematical Properties of the Ideal Circuit}\label{sec:math}

All claims concern the exact noiseless circuit under exact arithmetic. Full proofs appear in Appendix~\ref{appendix:proofs}.

\subsection{Permutation Invariance}\label{sec:math:invariance}

The encoding layer depends only on the multiset of vertex degrees; relabeling permutes the $R_X$ gates without changing the operators. The entangling layer consists of commuting $R_{ZZ}$ gates; relabeling permutes qubit positions without changing the operator family. The mixing layer is uniform and unaffected. Relabeling therefore permutes bitstring labels without changing the multiset of probabilities, and sorting removes this residual dependence.

\begin{theorem}[Permutation invariance]
\label{thm:invariance}
Let $G$ and $H$ be isomorphic simple undirected graphs. Then $\tilde{p}(G) = \tilde{p}(H)$.
\end{theorem}

\subsection{Injectivity and Completeness}\label{sec:math:injectivity}

The central question is whether distinct graphs can produce the same ideal output. We show they cannot, at one repetition and under an irrational-angle condition.

The argument rests on a reconstruction lemma.

\begin{lemma}[Cut reconstruction]
\label{lem:cut}
Let $G=([n],E)$ and $H=([n],E')$ be simple undirected graphs on the same labeled vertex set. If $\mathrm{cut}_G(s)=\mathrm{cut}_H(s)$ for all $s\in\{0,1\}^n$, then $E=E'$.
\end{lemma}

\begin{proof}
For $s\in\{0,1\}^n$,
\[
\mathrm{cut}_G(s) = \sum_{(u,v)\in E}(s_u + s_v - 2s_u s_v),
\]
a multilinear polynomial whose coefficient on $s_u s_v$ is $-2$ if $(u,v)\in E$ and $0$ otherwise. Multilinear polynomials on $\{0,1\}^n$ are uniquely determined by their values, so equal cut functions imply $E = E'$.
\end{proof}

Let $U_K = Z\,Y_K\,X_K$ denote the one-repetition circuit for graph $K$ (encoder $X_K$, entangler $Y_K$, mixer $Z$).

\begin{theorem}[One-repetition statevector injectivity]
\label{thm:injectivity}
Let $G$ and $H$ be simple undirected graphs on $[n]$. Assume $\theta_{\mathrm{enc}},\theta_{\mathrm{ent}},\theta_{\mathrm{mix}}\notin\pi\mathbb{Q}$, with $\theta_{\mathrm{enc}}<\pi$. If $U_G\lvert 0\rangle^{\otimes n} = U_H\lvert 0\rangle^{\otimes n}$, then $G = H$ as labeled graphs.
\end{theorem}

\emph{Proof sketch.} The common mixer reduces state equality to
post-entangler equality. Since the entangler is diagonal, taking
amplitude magnitudes pointwise isolates the encoder amplitudes ---
products of sines and cosines of degree-proportional angles.
Evaluating at the all-zeros and each Hamming-weight-one bitstring
recovers $\tan(\alpha_i)$ per vertex; injectivity of $\tan$ on
$[0,\pi/2)$ forces degree sequences to coincide. With encoders
equal, phase equality reduces to
$\exp(i\theta_{\mathrm{ent}}(\mathrm{cut}_G(s) -
\mathrm{cut}_H(s)))$ being constant in $s$. Because
$\theta_{\mathrm{ent}} \notin \pi\mathbb{Q}$, the only
integer-valued phase difference consistent with a constant unit
phase is zero; evaluating at $s = 0^n$ confirms the constant
vanishes, and Lemma~\ref{lem:cut} yields $G=H$. Full details
in Appendix~\ref{appendix:proofs:injectivity}.

\begin{corollary}[Complete embedding on isomorphism classes]
\label{cor:complete}
At one repetition and under the irrational-angle condition, $\tilde{p}(G) = \tilde{p}(H)$ if and only if $G \cong H$.
\end{corollary}

Two scope restrictions: (1) Theorem~\ref{thm:injectivity} concerns ideal statevectors under exact arithmetic, not finite-shot estimates; (2) the proof covers one repetition; the multi-repetition extension is conjectured in Appendix~\ref{appendix:proofs:injectivity} and supported empirically in Sections~\ref{sec:noiseless}--\ref{sec:hardware}. Hardware and noisy simulation necessarily use finite-precision realizations of the gate angles, so the theorem should be read as an ideal structural guarantee whose practical robustness is supported empirically rather than as a claim about exact hardware injectivity.
\subsection{From Ideal Distributions to Finite Shots}\label{sec:math:distinguishability}

In practice, QuIC is observed through finite-shot estimates. For $L_1$ distance, the triangle inequality gives
\begin{equation}
    \|\widehat{p}_G^{(N)} - \widehat{p}_H^{(N)}\|_1 \geq \|p_G - p_H\|_1 - \|\widehat{p}_G^{(N)} - p_G\|_1 - \|\widehat{p}_H^{(N)} - p_H\|_1,
\end{equation}
so distinguishability holds once estimation error falls below half the exact separation margin. This motivates both the shot-scaling analysis in Section~\ref{sec:noiseless} and the top-100 head truncation used throughout, since the distribution head carries most probability mass and is most reliably estimated.

\subsection{Global Encoding Structure}\label{sec:math:global}

\begin{figure}
    \centering
    \includegraphics[width=1\linewidth]{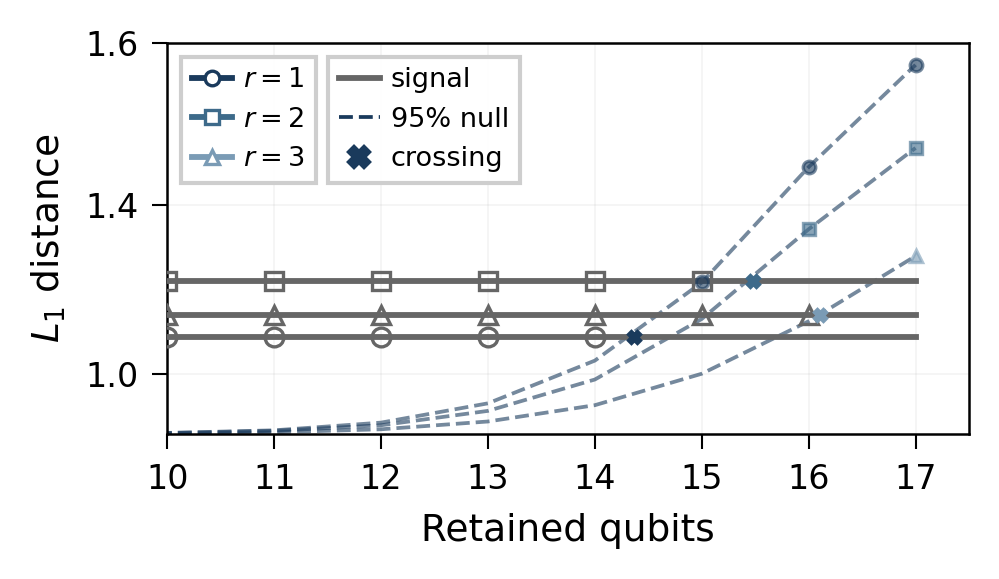}
    \caption{Global encoding structure of QuIC on broom$_{17,2}$ variants. The exact $L_1$ separation (solid) is constant regardless of how many path qubits are retained, confirming that structural information is encoded globally rather than propagated hop-by-hop. The sampling null (dashed) grows exponentially with retained qubits. Increasing repetitions compresses the null growth, extending the distinguishable range by 1--2 qubits per repetition, but reduces overall signal amplitude --- the same tradeoff observed in the repetition ablation (Section~\ref{sec:construction:grid}).}
    \label{fig:global}
\end{figure}

The completeness result establishes \emph{that} the ideal embedding
is injective; we now characterize \emph{how} the circuit distributes
structural information, which determines why practical truncation
remains effective despite exponential output dimension. We examine
a controlled example. A broom$_{17,2}$ graph consists of a star (centroid plus 2 pendants) whose centroid is joined to a path of length 14, totaling 17 nodes. We compare the base graph against a variant with one edge added between the two pendant nodes. The perturbation is confined to the star; the path portion is identical in both graphs.

For each cutoff $h \in \{0, \ldots, 14\}$, we trace out all path qubits beyond hop $h$ from the centroid and compute the $L_1$ distance of the resulting reduced distributions. At $h=0$ only the 3 pendant and centroid qubits are retained; at $h=14$ the full 17-qubit distribution is used (Figure~\ref{fig:global}).

The signal is effectively constant across all cutoffs: at $r=2$, the $L_1$ distance is 1.266 whether 3 or 17 qubits are retained. The perturbation's effect is not localized near the star --- it is present in the joint distribution at every scale. This is a direct consequence of the entangling layer: all $R_{ZZ}$ gates commute and act in a single step, encoding the full graph topology into the global phase structure simultaneously. Unlike iterative neighborhood refinement, where $k$ rounds of message passing access the $k$-hop neighborhood, every qubit's measurement outcome is conditioned on the entire graph from the first repetition. 

The sampling cost of accessing this information, however, grows exponentially with qubit count, as each additional qubit doubles the distribution dimension. The 95th-percentile null threshold --- estimated from same-circuit resampling at $2^{12}$ shots --- rises from 0.05 at 3 qubits to 1.48 at 17 qubits, eventually overtaking the fixed signal (Figure~\ref{fig:global}, crosses). Increasing repetitions compresses this null growth: $r=1$ loses distinguishability at approximately 15 retained qubits, $r=2$ at 16, and $r=3$ at 17. This is the same signal-versus-depth tradeoff observed in the repetition ablation (Section~\ref{sec:construction:grid}): additional repetitions reduce overall $L_1$ amplitude but improve noise resilience, extending the practical qubit range at a given shot budget.

In the tested regimes, the dominant practical bottleneck appears to be sampling cost rather than information reach. This directly motivates the sorted top-100 truncation (Section~\ref{sec:noiseless:distribution}): sorting collapses the exponentially growing distribution onto a fixed-length representation where signal dominates sampling noise.

\section{Noiseless Simulation Studies}\label{sec:noiseless}

All results use noiseless statevector simulation with canonical parameters ($\theta_{\mathrm{enc}} = 2.875$, $\theta_{\mathrm{ent}} = 2.0$, $\theta_{\mathrm{mix}} = 0.1$, $r = 2$) on ER random graphs unless stated otherwise.

\begin{figure}
    \centering
    \includegraphics[trim = {0.4cm, 0.42cm, 0cm, 0.26cm}, clip, width=\linewidth]{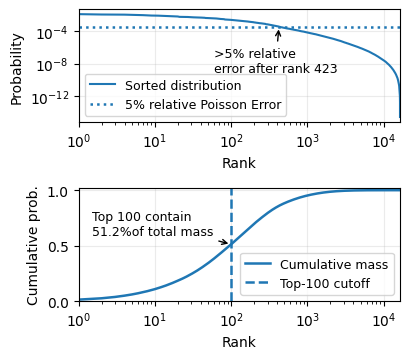}
    \caption{Sorted QuIC output for ER$(14,0.3)$, noiseless. (Top) Log--log decay; dotted line marks 5\% Poisson uncertainty. (Bottom) Cumulative mass; leading 100 entries capture 51.2\%.}
    \label{fig:zipf}
\end{figure}

\subsection{Distribution Structure and Head Truncation}\label{sec:noiseless:distribution}

The sorted output distribution $\tilde{p}(G)$ exhibits strongly skewed, approximately Zipf-like structure: a small number of high-probability outcomes concentrate the majority of the probability mass, with rapid decay across the tail (Figure~\ref{fig:zipf}). Graph-specific separation is concentrated in the head; tail entries are increasingly noise-dominated at practical shot counts.

This motivates the top-100 truncation used as the practical operating point in this study. The cutoff was chosen empirically on the basis of three observations: the leading 100 entries of the sorted distribution capture a large fraction of the total probability mass (51.2\% in the representative case of Figure~\ref{fig:zipf}), all 100 entries remain well above the 5\% relative Poisson uncertainty threshold in that example, and graph-specific separation is concentrated in this high-probability head. Additional top-$k$ checks at $k \in \{50,100,200,400\}$ showed that $k=50$ was insufficient to retain comparable stability, while gains beyond $k=100$ were modest relative to the added tail mass and sampling noise.

Across the tested ER pairs, this truncation preserved the study's operational separation criterion and retained the observed pairwise distinctions. We therefore use top-100 as an informed practical heuristic rather than as a theorem-derived or universal choice: its effectiveness is established only for the graph families, qubit ranges, and shot budgets studied here, whereas the completeness guarantee of Corollary~\ref{cor:complete} concerns the full sorted distribution.

\subsection{Finite-Shot Distinguishability}\label{sec:noiseless:shot_noise}

We measure operational separation using the $L_1$ distance between sorted distributions, comparing a \emph{null} baseline formed from 64 same-circuit subsamples against a \emph{signal} baseline formed from 64 subsampled estimates of two non-isomorphic graphs. Throughout the practical sections, a tested pair is said to satisfy  the study's separation criterion when the resulting $L_1$ z-score exceeds 3. We do not claim $z>3$ is a universal threshold for embedding discrimination; it is the conservative operational threshold adopted uniformly throughout this study.

Across tested ER pairs, $2^{12} = 4{,}096$ shots suffices for reliable separation in most cases (Figure~\ref{fig:finiteshot}), with $2^{15} = 32{,}768$ as a robust upper reference. All subsequent experiments use shot counts within this range, chosen to keep the practical QuIC embedding in a regime where the top-100 head entries are reliably estimable and graph-dependent separation remains visible above the null.

\begin{figure}
    \centering
    \includegraphics[trim = {0cm, 0.46cm, 0cm, 0.36cm},width=\linewidth]{finiteshot_2.png}
    \caption{Finite-shot separation for ER$(12,0.35)$ pair, top-100 truncated. Null and signal $L_1$ distances shown as mean $\pm 1\sigma$; dashed line is exact separation (0.0787). Vertical lines mark $2^{12}$ and $2^{15}$ subsampling levels.}
    \label{fig:finiteshot}
\end{figure}

\section{Noise-Model Simulation Studies}\label{sec:noisy}

All results use the canonical parameters simulated against a noise model derived from \texttt{ibm\_fez} calibration data (FakeFez). We identify effective mitigation strategies, then validate the embedding across all instance families tractable under classical noise-model simulation.

\subsection{Noise Model and Mitigation Study}\label{sec:noisy:model}

We evaluated six mitigation configurations (Table~\ref{tab:mitigation}) under FakeFez at $2^{12}$ shots on ER graphs ($n \in \{6,\ldots,15\}$, $p \in \{0.3,0.4,0.5\}$, degree-preserving rewiring pairs) and CFI pairs from P$_3$, P$_4$, C$_4$. Strategies tested include dynamical decoupling (DD)~\cite{viola1998dynamical,viola1999dynamical,pokharel2018demonstration,ezzell2023dynamical,tong2025empirical}, ZNE via circuit folding~\cite{temme2017error,kandala2019error,giurgica2020digital} and through Mitiq~\cite{larose2022mitiq}, Pauli twirling~\cite{wallman2016noise}, and measurement error mitigation. Separation was measured by $L_1$ z-score using the null/signal methodology of Section~\ref{sec:noiseless:shot_noise}.

DD (XX sequence) produced consistent improvement across all families and sizes and was adopted throughout; this aligns with broader superconducting-platform surveys~\cite{ezzell2023dynamical}. ZNE via folding showed minor gains but is intractable at hardware-relevant depths. Twirling, measurement mitigation, and library-based ZNE showed no consistent benefit.

\begin{table}[t]
    \centering
    \caption{Mitigation comparison on ER rewiring pairs and CFI pairs (P$_3$, P$_4$, C$_4$). *Mitiq ZNE: subset that completed within runtime.}
    \label{tab:mitigation}
    \begin{tabular}{lccc}
        \toprule
        \textbf{Strategy}   & \textbf{Mean $z$} & \textbf{$\Delta$} & \textbf{Cost} \\
        \midrule
        No mitigation       & 5.71              & ---               & $1\times$                 \\
        DD (XX)             & 5.83              & +2.1\%            & $1\times$ (fills idle)    \\
        DD (XpXm)           & 5.79              & +1.5\%            & $1\times$ (fills idle)    \\
        DD (XY4)            & 5.76              & +0.9\%            & $1\times$ (fills idle)    \\
        Pauli twirling      & 5.73              & +0.4\%            & ${\sim}1.1\times$         \\
        DD (XX) + Twirl     & 5.74              & +0.6\%            & ${\sim}1.1\times$         \\
        Mitiq ZNE*          & 5.76              & +0.9\%            & ${\sim}9\times$           \\
        \bottomrule
    \end{tabular}
\end{table}

\subsection{Validation Under Noise}\label{sec:noisy:validation}

We validated the embedding under FakeFez with DD at $2^{12}$ shots across four increasingly difficult instance families. All pairs used degree-preserving rewiring unless otherwise noted. 

\textbf{Exhaustive small graphs.} All non-isomorphic graphs with fewer
than 7 nodes satisfied the study's operational separation criterion,
with every pair exceeding twice the null baseline.

\textbf{Random graphs at scale.} ER graphs ($n \in \{7,\ldots,12\}$, $p \in \{0.25,\ldots,0.50\}$) and BA graphs ($n \in \{7,\ldots,12\}$, $m \in \{2,\ldots,8\}$), with up to $2^{15}$ samples and 2,000 rewiring attempts per graph. Across this campaign, approximately 38 million graphs were generated; exact global deduplication would therefore require an all-pairs isomorphism workload far beyond what was practical for the present study. In the rewiring experiments, each tested comparison graph was verified to be non-isomorphic to the graph it was paired against. All constructed pairs satisfied the study's operational separation criterion.

\textbf{Structured hard instances.} Three families probing distinctions invisible to degree sequences: chorded cycles ($C_n + e(0,k)$ vs.\ $C_n + e(0,j)$, $n \in \{12,15,18\}$), cycles with non-isomorphic inscribed triangles ($n \in \{12,15,18\}$), and ladder graphs with twisted rungs ($n \in \{12,15,18\}$). All tested pairs across these families satisfied the study's operational separation criterion.

\textbf{Classically difficult pairs.} We tested known hard instances from the strongly regular and regular graph literatures (Table~\ref{tab:srg}). The most demanding case is the Shrikhande graph versus the $4 \times 4$ rook's graph: both are strongly regular with parameters $(16,6,2,2)$, cospectral, and indistinguishable by 2-WL. QuIC separates them with $L_1 = 0.1710$ and $z = 5.16$. The remaining pairs --- the Petersen graph, cube $Q_3$, and $L(K_{2,4})$, each compared against a degree-matched regular counterpart --- are indistinguishable by 1-WL and were all separated with comparable margins. These results complement the CFI experiments: where CFI pairs probe the WL hierarchy through constructed gadgets, the instances in Table~\ref{tab:srg} are naturally occurring graphs whose difficulty arises from spectral and structural coincidences.

\begin{table}[t]
    \centering
    \caption{Classically difficult graph pairs under FakeFez
    noise model with DD at $2^{12}$ shots.}
    \label{tab:srg}
    \setlength{\tabcolsep}{5pt}
    \renewcommand{\arraystretch}{1.18}
    \begin{tabular}{lcccc}
        \toprule
        \textbf{Pair} & \textbf{$n$} & \textbf{WL failure}
            & \textbf{$L_1$} & \textbf{$z$} \\
        \midrule
        Shrikhande vs.\ Rook$_{4 \times 4}$     & 16 & 2-WL & 0.1710 & 5.16 \\
        Petersen vs.\ Reg(3,10)                 & 10 & 1-WL & 0.2314 & 7.55 \\
        $Q_3$ vs.\ Reg(3,8)                     & 8  & 1-WL & 0.2194 & 6.91 \\
        $L(K_{2,4})$ vs.\ Reg(4,8)              & 8  & 1-WL & 0.1116 & 5.54 \\
        \bottomrule
    \end{tabular}
\end{table}

\subsection{CFI Validation Under Noise}\label{sec:noisy:cfi_sim}

CFI pairs are constructed from a base graph $G$ by replacing each
vertex with a gadget and each edge with a connection pattern,
producing an untwisted/twisted pair that differs by a local gadget
modification. The resulting graphs are indistinguishable by any
fixed-$k$ WL refinement~\cite{cai1992optimal}. The vertex count is
\begin{equation}
    |V_{\mathrm{CFI}}| = \sum_{v \in V(G)}
    \left(2^{\deg(v)-1} + 2\,\deg(v)\right),
    \label{eq:cfi_size}
\end{equation}
which grows exponentially in vertex degree, explaining why dense
base graphs produce the largest circuits and encounter the depth
limit first (Section~\ref{sec:hardware:wall}). Full construction
details appear in Appendix~\ref{appendix:CFI_construction}.

Four pairs tractable under classical noise-model simulation were
tested: CFI($P_4$), CFI($P_5$), CFI($P_6$), and CFI($K_3$). All
four were successfully separated. CFI($K_3$) achieved
$z = 6.93$, confirming separation of a genuine WL-hard instance
under realistic noise. These results justified the hardware
investment for larger CFI families beyond the classical simulation
budget.

\section{Hardware Study}\label{sec:hardware}

All hardware experiments used \texttt{ibm\_fez}, a 156-qubit IBM Heron r2 processor with heavy-hex connectivity and tunable couplers~\cite{abughanem2025ibm}, executed via Qiskit~\cite{qiskit2024}. This device generation has demonstrated execution of circuits with thousands of two-qubit gates under noise-aware mitigation~\cite{kim2023evidence}. All runs used $2^{15} = 32{,}768$ shots; separation was assessed by $L_1$ z-score ($z > 3$ threshold). All hardware experiments evaluate practical realizations of the QuIC embedding under finite-shot sampling, transpilation, and device noise. The hardware study includes both the one-repetition circuit covered by the injectivity theorem and the empirically stronger two-repetition circuit, allowing direct comparison between the theorem-backed repetition regime and the practically preferred higher-repetition setting. Formal completeness itself applies only to the ideal exact one-repetition object defined in Section~\ref{sec:math}.

\subsection{Depth Limit}\label{sec:hardware:wall}

Initial hardware experiments on CFI pairs at 30--32 qubits (P$_6$, C$_5$, Pan$_5$, K$_4$-$e$) revealed two practical execution issues that shaped the remainder of the study. First, logically equivalent transpilation outcomes could differ substantially in observed hardware performance: circuits produced under \texttt{optimization\_level=3} often exhibited weaker separation than barrier-stabilized versions of the same logical circuit despite no identified logical error in the transpiled unitaries. We do not claim to explain that compiler-level mechanism here; instead, we treat barrier-stabilized transpilation as part of the reported hardware execution protocol and characterize the effect empirically in Section~\ref{sec:hardware:transpilation}.

\begin{figure}
    \centering
    \includegraphics[width=1\linewidth]{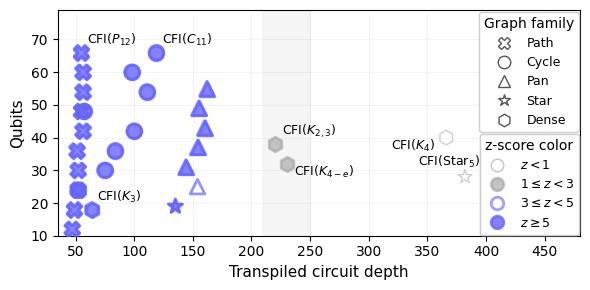}
    \caption{Hardware feasibility map on \texttt{ibm\_fez}. Each point is one CFI instance by transpiled depth and qubit count; shape denotes base-graph family, color denotes z-score. Shaded band: depth limit transition region (depths 210--250).}
    \label{fig:thermalization}
\end{figure}

Second, circuits below approximately 210--250 layers depth produced meaningful separation, while deeper circuits collapsed toward a noise-dominated regime regardless of graph family. We term this the \emph{depth limit} (Figure~\ref{fig:thermalization}). This depth boundary motivates the paired hardware comparison of reps$=2$ and reps$=1$ reported later: the additional repetition often improves empirical separation, but it also increases circuit depth enough to matter near the device noise horizon.

\subsection{Transpilation Ablation}\label{sec:hardware:transpilation}

\begin{figure*}
    \centering
    \includegraphics[width=1\linewidth]{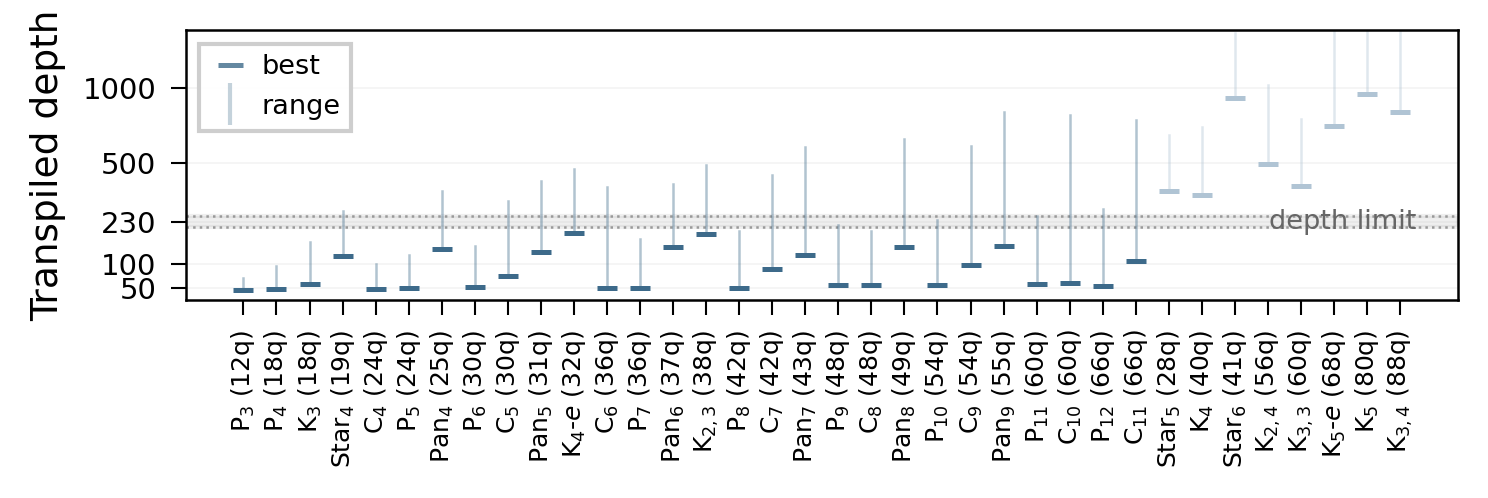}
    \caption{Transpiled depth across 200 transpilations (20 edge orderings $\times$ 10 seeds) per CFI family on \texttt{ibm\_fez}, sorted by qubit count within feasible and infeasible groups. Path and cycle families remain well below the depth limit regardless of qubit count, while dense and star-like families exceed it even at modest sizes --- feasibility is governed by base-graph density rather than circuit width.}
    \label{fig:transpilation}
\end{figure*}

We transpiled 14,800 circuits across all 37 CFI base graph pairs (20 edge orderings $\times$ 10 random seeds $\times$ 2 twisted/untwisted) targeting \texttt{ibm\_fez} (Figure~\ref{fig:transpilation}). The primary finding is that transpiled depth is governed by the density of the base graph, not by qubit count. Path families achieve best depths of 46--56 layers from 12 to 66 qubits; cycle families range from 48 to 106. Both remain well below the depth limit across the full qubit range. By contrast, Star$_5$ (28 qubits) reaches a best depth of 360, and K$_3$ (18 qubits) reaches 56 --- a 6$\times$ difference at half the qubit count, driven entirely by edge density.

Within each family, transpiled depth varies by up to an order of magnitude depending on edge ordering and random seed: CFI(C$_6$) ranged from 50 to 377 (7.5$\times$), CFI(C$_{10}$) from 59 to 800 (13.5$\times$), and CFI(Pan$_9$) from 147 to 820 (5.6$\times$). This variance makes single-transpilation execution unstable for this circuit family on current hardware. Accordingly, the reported hardware results use the best-depth outcome from the full multi-seed sweep as part of the execution protocol studied here. We emphasize that this sweep is not part of the mathematical method itself, but of the practical procedure required to execute these graph-dependent circuits on \texttt{ibm\_fez}. Full seed and ordering data are provided in Appendix~\ref{appendix:transpilation_depth}. 

Dense families --- complete graphs, complete bipartite graphs, and stars --- exceed the depth limit even at their smallest instances. K$_4$ (40 qubits) has a best depth of 340; Star$_6$ (41 qubits) reaches 922. These families are infeasible under the canonical reps$=2$ circuit regardless of transpilation effort, motivating the paired hardware comparison with reps$=1$ in Section~\ref{sec:hardware:degradation}. The one-repetition circuit both restores the repetition regime covered by the current theory and reduces depth in boundary cases.

A secondary finding is that the shallowest transpilation is not always the best-performing on hardware. Some lower-depth outcomes under \texttt{optimization\_level=3} exhibited weaker separation than barrier-stabilized alternatives despite no identified logical error. We therefore scope all reported hardware conclusions to the barrier-stabilized execution protocol used here. The underlying compiler-level mechanism remains unresolved and is left for future systems investigation.

\subsection{Hardware-Adapted Execution}\label{sec:hardware:degradation}

The parameter ablation of Section~\ref{sec:construction:grid} identifies reps$=2$ as the preferred practical operating point in simulation. On hardware, however, the additional repetition must be balanced against the depth cost it introduces. To characterize this tradeoff directly, we report paired hardware runs at both reps$=2$ and reps$=1$ wherever feasible. The one-repetition circuit is \begin{equation} 
    U'(G) = U_{\mathrm{mix}}\,U_{\mathrm{ent}}\,U_{\mathrm{enc}}.
\end{equation}
This is the repetition regime covered by the injectivity theorem, so its hardware execution provides the closest practical realization of the proven setting, though under finite-shot estimation, finite precision, transpilation, and device noise rather than the ideal exact regime.

The paired study shows that reps$=2$ remains the stronger hardware setting in most tested instances, consistent with the simulation ablations. At the same time, the shallower reps$=1$ circuits recover distinguishability in some depth-limited cases, especially among denser families near the hardware boundary. Repetition count on hardware is therefore best understood as a depth-sensitive tradeoff: reps$=2$ is usually preferable when feasible, while reps$=1$ provides both the theorem-covered repetition regime and a useful lower-depth alternative in boundary cases.

\subsection{Experimental Protocol}\label{sec:hardware:protocol}

All runs used $2^{15}$ shots, subsampled at $2^{12}$ for statistical analysis. DD (XX sequence) was applied via \texttt{PadDynamicalDecoupling}, consistent with the noise-model study. Separation was assessed by $L_1$ z-score:
\begin{equation}
    z = \frac{\mu_{\text{signal}} - \mu_{\text{null}}}
             {\sigma_{\text{pooled}}},
\end{equation}
where the null is formed from same-circuit subsample comparisons and the signal from untwisted/twisted pair comparisons; $z > 3$ is the threshold throughout. Path and cycle families used manual layouts excluding the worst-performing 30\% of two-qubit gate positions based on device calibration data --- selected empirically as the lowest exclusion level that consistently yielded feasible-depth mappings; all others used best-seed transpiled layouts from the ablation study.

Throughout the hardware section, a tested pair is counted as successfully separated when its signal distribution exceeds the same-circuit null under this protocol with $z>3$. This operational criterion is the sense in which we use terms such as ``separated'' in the practical experiments.

\subsection{Hardware Results}\label{sec:hardware:extended}

Table~\ref{tab:hardware_results} reports paired hardware results at reps$=2$ and reps$=1$. The dominant pattern is that reps$=2$ remains the stronger setting in most tested instances. This is especially clear across the manually laid out path families P$_7$--P$_{12}$, most cycle families C$_7$--C$_{11}$, and several smaller structured families such as C$_5$ and Pan$_5$--Pan$_8$, where the additional repetition continues to translate into stronger observed separation.

The paired reps$=1$ runs nonetheless reveal an important boundary effect. On some denser or more depth-stressed circuits, reducing repetition lowers transpiled depth enough to recover distinguishability lost at reps$=2$. This is most visible in K$_4$-$e$, which rises from $z=1.64$ at depths $230/230$ under reps$=2$ to $z=3.81$ at depths $170/148$ under reps$=1$, and in K$_{2,3}$, which rises from $z=2.15$ at $220/220$ to $z=3.27$ at $188/169$. A smaller but similar gain appears for P$_6$.  These comparisons refine rather than overturn the operating-point selection from simulation. Reps$=2$ remains the preferred practical hardware setting overall, but the paired study shows that this preference weakens near the device depth boundary, where the theorem-covered one-repetition circuit can become advantageous in practice. The largest remaining hard case is CFI(K$_{2,4}$), which remains below threshold even at reps$=1$ ($z=1.07$ at depths $253/251$), marking the current practical boundary of the method. 

\begin{table}[t]
    \centering
    \caption{Paired hardware results on \texttt{ibm\_fez} at reps$=2$ and reps$=1$. Depths: untwisted/twisted. $^\dagger$Manual layouts; $^\ddagger$reps$\,=1$ only.}    \label{tab:hardware_results}
    \setlength{\tabcolsep}{7pt}
    \renewcommand{\arraystretch}{1.18}
    \begin{tabular}{lc|cc|cc}
        \toprule
                            &                  & \multicolumn{2}{c}{reps = 2}   & \multicolumn{2}{c}{reps = 1} \\               
        \textbf{Base}       & $\mathbf{|V(G)|}$& \textbf{Depth} &               & \textbf{Depth}&              \\
        \textbf{graph}      & \textbf{Qubits}  & \textbf{(untw/tw)}& \textbf{$z$}& \textbf{(untw/tw)}& \textbf{$z$} \\
        \midrule
        P$_6$               & 30               & 148/148        & 5.08          &  119/108          &  8.11         \\
        C$_5$               & 30               & 165/165        & 18.07         &  112/99           &  7.98         \\
        Pan$_5$             & 31               & 144/157        & 5.88          &  97/83            &  5.51             \\
        K$_4$-$e$           & 32               & 230/230        & 1.64          &  170/148          &  3.81             \\
        Pan$_6$             & 37               & 154/161        & 5.26          &  111/112          &  5.49             \\
        Pan$_7$             & 43               & 160/159        & 5.26          &  119/131          &  3.98             \\
        Pan$_8$             & 49               & 155/173        & 5.14          &  138/120          &  4.05             \\
        K$_{2,3}$           & 38               & 220/220        & 2.15          &  188/169          &  3.27             \\
        
        P$_7$$^\dagger$     & 36               & 156/156        & 7.33          & 133/111           & 6.50              \\
        P$_8$$^\dagger$     & 42               & 170/166        & 7.09          & 137/120           & 4.46              \\
        P$_9$$^\dagger$     & 48               & 182/183        & 7.34          & 151/167           & 3.73              \\
        P$_{10}$$^\dagger$  & 54               & 196/191        & 7.94          & 191/175           & 4.08              \\
        P$_{11}$$^\dagger$  & 60               & 203/212        & 9.65          & 182/169           & 5.04              \\
        P$_{12}$$^\dagger$  & 66               & 218/215        & 8.14          & 201/194           & 4.25              \\
        
        C$_6$$^\dagger$     & 36               & 184/192          & 6.44        & 151/123           & 6.51              \\
        C$_7$$^\dagger$     & 42               & 200/212        & 6.89          & 177/159           & 6.09              \\
        C$_8$$^\dagger$     & 48               & 157/160          & 7.05        & 169/174           & 4.14              \\
        C$_9$$^\dagger$     & 54               & 211/223        & 8.07          & 180/197           & 5.36              \\
        C$_{10}$$^\dagger$  & 60               & 198/211         & 9.34         & 196/202           & 4.98              \\
        C$_{11}$$^\dagger$  & 66               & 219/231        & 7.03          & 209/199           & 4.68              \\
        
        K$_{3,3}$$^\ddagger$& 60               &         &           &         215/218          &     3.72          \\
        K$_{2,4}$$^\ddagger$& 56               &         &           &    253/251               &     1.07          \\
        \bottomrule
    \end{tabular}
\end{table}

\subsection{Marginal Cases}\label{sec:hardware:marginal}

Most hardware-tested instances favor reps$=2$, but the paired repetition study identifies a small set of depth-limited cases where reps$=1$ is beneficial. Under reps$=2$, CFI(K$_4$-$e$) achieves $z=1.64$ at $2^{12}$ shots but rises to 2.34 at $2^{13}$ and 3.17 at $2^{14}$ from the same $2^{15}$ run, indicating that the reps$=2$ separation margin is real but narrow. Under reps$=1$, the same instance rises to $z=3.81$ with substantially reduced depth. Likewise, CFI(K$_{2,3}$) rises from $z=2.15$ under reps$=2$ to $z=3.27$ under reps$=1$. In both cases, the evidence indicates that the reps$=2$ degradation is primarily depth-driven rather than structural.

CFI(K$_{2,4}$) was intractable at reps = 2, and, even at reps = 1, was near the depth limit and was not able to meet our separation criteria with a $z \approx 1$. This indicates a genuine depth-limit failure rather than insufficient shots.

\section{Discussion}\label{sec:discussion}

Formal completeness applies only to the ideal QuIC embedding under exact arithmetic: the full sorted output distribution of the one-repetition circuit. All experiments instead evaluate practical QuIC realizations under finite-shot estimation and, on hardware, under transpilation, execution heuristics, and device noise. The role of the ideal theory in this paper is therefore motivational and structural: it identifies a graph-to-circuit map with strong properties in the exact one-repetition regime, then guides the construction of practical embeddings whose robustness must be evaluated empirically rather than assumed from the theorem.

Within that practical regime, fixed-length head truncation serves as an informed compression strategy rather than as a universal or theorem-backed choice: the distribution head concentrates much of the probability mass and remains the most reliably estimable portion at practical shot counts, while the tail is increasingly noise-dominated (Section~\ref{sec:noiseless:distribution}). The hardware study correspondingly includes the theorem-covered one-repetition circuit as the closest practical realization of the proven regime, alongside higher-repetition variants that often perform better empirically but remain outside the current proof once $r>1$.

The global encoding analysis (Section~\ref{sec:math:global}) clarifies why these concessions work. Because the entangling layer acts in a single commuting step, structural information is encoded globally rather than propagated through local neighborhoods --- the discriminative signal is present at every scale, and the practical limit is sampling cost, not information reach. This is also why the embedding is not naturally situated within the WL hierarchy, rather than being a quantum analogue of it: there is no iterative refinement whose depth determines expressiveness. The top-100 truncation succeeds precisely because sorting collapses the exponentially growing distribution onto a fixed-length representation where signal dominates noise, and additional repetitions extend this regime modestly by compressing the null growth rate.

The depth limit near 210--250 layers on \texttt{ibm\_fez} provides a concrete engineering constraint: logical expressiveness is insufficient unless compilation and execution remain below the device noise horizon. The paired hardware study shows that reps$=2$ remains the stronger practical setting in most tested instances, but that this advantage can weaken near the boundary where added repetition pushes circuits too close to the noise horizon. CFI(K$_4$-$e$) and CFI(K$_{2,3}$) illustrate this effect: both fall below threshold at reps$=2$ but recover above threshold at reps$=1$. By contrast, CFI(K$_{2,4}$) remains below threshold even at reps$=1$, marking the current boundary. A related finding is transpilation sensitivity: barrier-stabilized circuits consistently outperformed some lower-depth \texttt{optimization\_level=3} alternatives despite no identified logical error. The mechanism is unidentified; for large graph-dependent circuits, nominal depth reduction alone does not predict hardware performance.

The CFI and strongly regular graph experiments demonstrate empirical separation on instances designed to defeat fixed-$k$ WL refinement. We do not claim a formal separation from the WL hierarchy. 

\textbf{Limitations.} Injectivity is proven only for the one-repetition ideal exact setting; the multi-repetition extension remains an open theoretical problem. The hardware study includes the one-repetition circuit covered by the theorem, but only as a practical realization under finite-shot estimation, transpilation, execution heuristics, and device noise rather than as the ideal exact object itself. In the practical sections, empirical success means only that the tested graph pairs satisfied the study's operational separation criterion; it should not be read as a completeness guarantee for noisy, truncated, or repeated circuits. Hardware results are device- and calibration-specific, and the reported performance is conditional on the execution protocol used here. The embedding addresses unlabeled topological structure only.

\textbf{Future work.} On the theory side, extending injectivity to the full repeated circuit is the most important open problem. On the systems side, the transpilation-sensitivity effect deserves deeper investigation. On the application side, the most immediate extension is attributed-graph encoding via feature reuploading for molecular and chemical graph data. Beyond that, kernelization of the QuIC embedding for supervised classification on standard benchmarks (MUTAG, PTC, PROTEINS) and integration as an encoder layer within quantum graph neural network architectures are natural next steps.

\section{Conclusion}\label{sec:conclusion}

We introduced QuIC, a training-free quantum graph embedding whose ideal one-repetition sorted distribution is complete on isomorphism classes under the stated irrational-angle condition. We then used that ideal analysis to motivate a practical embedding pipeline and studied its behavior under finite-shot estimation, truncation, noise, transpilation, and hardware execution. Under noise-model simulation, all tested pairs in the reported exhaustive, random, structured, and CFI suites satisfied the study's operational separation criterion. On IBM Heron hardware, 14,800 transpiled circuits across 37 CFI graph families demonstrated empirical separation up to 66 qubits for the tested families, identified a device-dependent depth limit near 210--250 layers, and characterized the current practical boundary of the method under the reported execution protocol.

The paired hardware study further shows that the theorem-covered one-repetition circuit and the empirically stronger two-repetition circuit play distinct but complementary roles under practical constraints: reps$=2$ usually provides the strongest separation, while reps$=1$ recovers some depth-limited cases and provides the closest practical realization of the proven repetition regime. The main contribution is therefore not a claim that the practical hardware object is complete, but a combined result: an ideal theoretical analysis that motivates the design, together with a large-scale empirical study of how far that design remains useful under realistic execution constraints.

\bibliographystyle{IEEEtran}
\bibliography{references}

\appendices

\newpage
\section{CFI Construction Details and Full Graph Family List}
\label{appendix:CFI_construction}

This appendix gives the explicit CFI gadget construction used
throughout the paper, following Cai, F\"{u}rer, and
Immerman~\cite{cai1992optimal}, and lists the 37 base graphs used in
the hardware transpilation sweep and hardware execution study
(Sections~\ref{sec:hardware:transpilation}
and~\ref{sec:hardware:extended}).

\subsection{Gadget Construction}
\label{appendix:CFI_construction:gadget}

Let $G = (V, E)$ be a simple connected base graph. For each vertex
$v \in V$ with $\deg(v) = d$, the CFI construction introduces a local
gadget consisting of two groups of vertices:

\begin{enumerate}
    \item An \emph{edge-vertex block} $E_v$ of size $2d$: for each
    edge $e = \{v, u\} \in E$ incident to $v$, two vertices
    $e_v^{(0)}$ and $e_v^{(1)}$.
    \item An \emph{inner-vertex block} $M_v$ of size $2^{d-1}$: one
    vertex for each binary string
    $s \in \{0,1\}^d$ of \emph{even} Hamming weight.
\end{enumerate}

Inner vertices are wired to edge vertices according to the bits of
their labels: inner vertex $s = (s_1, \ldots, s_d)$ is connected to
$e_i^{(s_i)}$ for each incident edge $e_i$, where the edges incident
to $v$ are ordered arbitrarily. The restriction to even-weight strings
is the construction's key structural feature: it produces a gadget
whose automorphism group acts as the even subgroup of
$(\mathbb{Z}/2\mathbb{Z})^d$, permitting simultaneous bit-flips on any
even number of incident edges but forbidding odd flips.

Each base edge $e = \{u, v\} \in E$ is then replaced by a two-edge
bridge $e_u^{(b)} - e_v^{(b)}$ for each $b \in \{0, 1\}$, producing
two parallel bridges connecting the two endpoint gadgets. The total
vertex count of the CFI graph follows by summing gadget sizes:
\[
|V_{\mathrm{CFI}}| = \sum_{v \in V(G)}\bigl(2^{\deg(v) - 1} + 2\,\deg(v)\bigr),
\]
which is Equation~(\ref{eq:cfi_size}).

\subsection{Twisting}
\label{appendix:CFI_construction:twist}

The untwisted CFI graph, denoted $\mathrm{CFI}(G)$, is obtained by
applying the construction above directly. The twisted variant,
denoted $\widetilde{\mathrm{CFI}}(G)$, is obtained by selecting a
single base edge $e^* = \{u, v\} \in E$ and swapping its bridges:
$e_u^{(0)}$ is connected to $e_v^{(1)}$ and $e_u^{(1)}$ to $e_v^{(0)}$,
while every other base edge retains its parallel bridge structure.
The choice of twist edge $e^*$ does not affect isomorphism type:
swapping any odd number of bridges yields the same isomorphism class,
and swapping any even number yields the untwisted graph.

The central property exploited in this paper is that
$\mathrm{CFI}(G)$ and $\widetilde{\mathrm{CFI}}(G)$ are non-isomorphic
when $G$ is connected and has minimum degree at least two, yet are
indistinguishable by the $k$-dimensional Weisfeiler--Leman
refinement for any fixed $k$ when $G$ has sufficiently large treewidth
relative to $k$~\cite{cai1992optimal}. CFI families therefore provide
canonical stress tests for any graph representation bounded above by
fixed-$k$ WL expressiveness.

\subsection{Base Graph Families}
\label{appendix:CFI_construction:families}

The hardware transpilation sweep and execution study used 37 base
graphs drawn from six structural families, chosen to span a range of
vertex counts, degree sequences, and treewidths while keeping the
resulting CFI qubit counts tractable on
\texttt{ibm\_fez}. Table~\ref{tab:cfi_families} lists the complete
set, grouped by family; qubit counts reflect the full CFI gadget
expansion via Equation~(\ref{eq:cfi_size}).

\begin{table}[ht]
    \centering
    \caption{CFI base graphs used in the hardware study, grouped by
    family. ``Qubits'' is the vertex count of the resulting CFI graph
    (untwisted and twisted are the same size).}
    \label{tab:cfi_families}
    \setlength{\tabcolsep}{5pt}
    \renewcommand{\arraystretch}{1.15}
    \begin{tabular}{llcc}
        \hline
        \textbf{Family} & \textbf{Base graph} & $|V(G)|$ & \textbf{Qubits} \\
        \hline
        Path       & P$_3$        & 3   & 12 \\
                   & P$_4$        & 4   & 18 \\
                   & P$_5$        & 5   & 24 \\
                   & P$_6$        & 6   & 30 \\
                   & P$_7$        & 7   & 36 \\
                   & P$_8$        & 8   & 42 \\
                   & P$_9$        & 9   & 48 \\
                   & P$_{10}$     & 10  & 54 \\
                   & P$_{11}$     & 11  & 60 \\
                   & P$_{12}$     & 12  & 66 \\
        \hline
        Cycle      & C$_4$        & 4   & 24 \\
                   & C$_5$        & 5   & 30 \\
                   & C$_6$        & 6   & 36 \\
                   & C$_7$        & 7   & 42 \\
                   & C$_8$        & 8   & 48 \\
                   & C$_9$        & 9   & 54 \\
                   & C$_{10}$     & 10  & 60 \\
                   & C$_{11}$     & 11  & 66 \\
        \hline
        Pan        & Pan$_4$      & 5   & 25 \\
                   & Pan$_5$      & 6   & 31 \\
                   & Pan$_6$      & 7   & 37 \\
                   & Pan$_7$      & 8   & 43 \\
                   & Pan$_8$      & 9   & 49 \\
                   & Pan$_9$      & 10  & 55 \\
        \hline
        Complete   & K$_3$        & 3   & 18 \\
                   & K$_4$-$e$    & 4   & 32 \\
                   & K$_4$        & 4   & 40 \\
                   & K$_5$-$e$    & 5   & 68 \\
                   & K$_5$        & 5   & 80 \\
        \hline
        Star       & Star$_4$     & 5   & 19 \\
                   & Star$_5$     & 6   & 28 \\
                   & Star$_6$     & 7   & 41 \\
                   & Star$_7$     & 8   & 62 \\
        \hline
        Bipartite  & K$_{2,3}$    & 5   & 38 \\
                   & K$_{2,4}$    & 6   & 56 \\
                   & K$_{3,3}$    & 6   & 60 \\
                   & K$_{3,4}$    & 7   & 88 \\
        \hline
    \end{tabular}
\end{table}

\newpage
\section{Parameter Sweep Details}
\label{appendix:sweeps}

Table~\ref{tab:enc_sweep} in the main text reports the encoding-angle
sweep. This appendix provides the companion sweeps for the entangling
angle, mixing angle, and repetition count.

\subsection{Entangling and Mixing Angle Sweeps}
\label{appendix:sweeps:ent_mix}

The entangling- and mixing-angle sweeps were run as a joint refined
grid over $\theta_{\mathrm{enc}} \in \{2.500, 2.625, \ldots, 3.500\}$,
$\theta_{\mathrm{ent}} \in \{1.750, 1.875, \ldots, 2.250\}$, and
$\theta_{\mathrm{mix}} \in \{0.00, 0.05, 0.10, 0.15, 0.20\}$ on
CFI(P$_4$) and CFI(Star$_4$) at $r = 2$ and 4{,}096 shots, with
$z$-scores computed from 64 same-circuit null subsamples and 64
pairwise non-isomorphic signal subsamples. Marginal entries in
Table~\ref{tab:ent_sweep} below average over the non-swept
parameters within this grid; the corresponding mixing-angle sweep
(Table~\ref{tab:mix_sweep}) is characterized in two regimes.

\begin{table}[ht]
    \centering
    \caption{Entangling-angle sweep, marginal over
    $\theta_{\mathrm{enc}}$ and $\theta_{\mathrm{mix}}$.}
    \label{tab:ent_sweep}
    \setlength{\tabcolsep}{7pt}
    \renewcommand{\arraystretch}{1.18}
    \begin{tabular}{cccc}
        \hline
        $\theta_{\mathrm{ent}}$ & $\bar{z}$ (mean) & $z_{\min}$ (worst) & Status \\
        \hline
        1.750               & 3.43          & 2.29          & Flat          \\
        1.875               & 3.41          & 2.14          & Flat          \\
        \textbf{2.000}      & \textbf{3.42} & \textbf{2.22} & \textbf{Selected} \\
        2.125               & 3.43          & 2.01          & Flat          \\
        2.250               & 3.42          & 2.32          & Flat          \\
        \hline
    \end{tabular}
\end{table}

Mean $z$-score varies by less than $0.02$ across the interval, so
$\theta_{\mathrm{ent}}$ is effectively a flat direction within the
swept range. $\theta_{\mathrm{ent}} = 2.0$ is adopted as the
canonical value.

\begin{table}[ht]
    \centering
    \caption{Mixing-angle sweep. Top panel: refined sweep, marginal
    over $\theta_{\mathrm{enc}} \in [2.5, 3.5]$ and
    $\theta_{\mathrm{ent}} \in [1.75, 2.25]$ on CFI(P$_4$) and
    CFI(Star$_4$). Bottom panel: coarse sweep at the fixed slice
    $\theta_{\mathrm{enc}} = 3.0$, $\theta_{\mathrm{ent}} = 2.0$
    across four CFI pairs (P$_3$, K$_3$, P$_4$, Star$_4$),
    showing degradation at large mixing angles. Both panels
    use 4{,}096 shots, $r=2$.}
    \label{tab:mix_sweep}
    \setlength{\tabcolsep}{7pt}
    \renewcommand{\arraystretch}{1.18}
    \begin{tabular}{cccc}
        \hline
        $\theta_{\mathrm{mix}}$ & $\bar{z}$ (mean) & $z_{\min}$ (worst) & Status \\
        \hline
        \multicolumn{4}{l}{\textit{Refined sweep (near-optimum)}}             \\
        0.00                & 3.29          & 2.01          & No mixer           \\
        0.05                & 3.45          & 2.37          & Improved           \\
        \textbf{0.10}       & \textbf{3.52} & \textbf{2.44} & \textbf{Selected}  \\
        0.15                & 3.42          & 2.44          & Stable             \\
        0.20                & 3.42          & 2.50          & Stable             \\
        \hline
        \multicolumn{4}{l}{\textit{Coarse sweep at $\theta_{\mathrm{enc}}=3.0$, $\theta_{\mathrm{ent}}=2.0$}} \\
        0.10                & 3.68          & 1.86          & Near canonical     \\
        0.30                & 3.43          & 2.78          & Acceptable         \\
        0.50                & 3.06          & 2.38          & Weakening          \\
        0.70                & 2.81          & 2.46          & Near threshold     \\
        0.90                & 1.71          & 0.79          & Degraded           \\
        1.10                & $-$0.31       & $-$1.64       & Failed             \\
        \hline
    \end{tabular}
\end{table}

A small, nonzero mixer improves both mean and worst-case separation
over the $\theta_{\mathrm{mix}} = 0$ baseline, with
$\theta_{\mathrm{mix}} = 0.10$ giving the best mean $z$ in the
refined range; worst-case $z$ continues to rise modestly through
$0.20$ but without further mean gain. Separation degrades at larger
mixer angles outside this range (see Section~\ref{sec:construction:grid}).

\subsection{Repetition Sweep}
\label{appendix:sweeps:reps}

The repetition sweep was run under exact statevector simulation on
ER, BA, and SRG motif collections drawn from the group1 dataset
(subgraph sizes $k \in \{8,10,12,16\}$), with
$r \in \{1,2,3,4\}$. Pairwise total-variation distances between
non-isomorphic motif fingerprints were averaged within each family.
Table~\ref{tab:reps_sweep} reports the comparison at the
representative operating point of the sweep.

\begin{table}[ht]
    \centering
    \caption{Repetition sweep: mean pairwise TV distance between
    non-isomorphic fingerprints by family, at the representative
    sweep operating point. Statevector simulation, no shot noise.}
    \label{tab:reps_sweep}
    \setlength{\tabcolsep}{7pt}
    \renewcommand{\arraystretch}{1.18}
    \begin{tabular}{ccccc}
        \hline
        $r$ & ER & BA & SRG & Status \\
        \hline
        1                   & 0.301          & 0.240          & 0.312          & Baseline                  \\
        \textbf{2}          & \textbf{0.321} & \textbf{0.251} & \textbf{0.323} & \textbf{Selected}         \\
        3                   & 0.310          & 0.239          & 0.320          & Comparable to $r=1$       \\
        4                   & 0.295          & 0.226          & 0.313          & Degraded (ER, BA)         \\
        \hline
    \end{tabular}
\end{table}

Across all three families, $r=2$ gives the largest mean TV
separation. $r=3$ is comparable to $r=1$ on ER and BA and slightly
below $r=2$ on SRG; $r=4$ drops below $r=1$ on ER and BA, consistent
with additional rounds driving the output toward uniformity.

\newpage

\section{Reproducibility and Software Setup}
\label{appendix:reproducibility}

This appendix documents the software environment, execution protocol,
and statistical methodology used throughout the paper, with the aim of
making the reported results independently reproducible subject to the
device- and calibration-specific caveats discussed in
Section~\ref{sec:discussion}.

\subsection{Software Stack}
\label{appendix:reproducibility:software}

All circuits were constructed and executed using the Qiskit ecosystem:
\texttt{qiskit} 2.4.0, \texttt{qiskit-aer} 0.17.2, and
\texttt{qiskit-ibm-runtime} 0.46.1. Noiseless simulation used
\texttt{qiskit\_aer}'s \texttt{SamplerV2} primitive. Noise-model
simulation used \texttt{FakeFez} from
\texttt{qiskit-ibm-runtime.fake\_provider}, which constructs a
\texttt{NoiseModel} from the \texttt{ibm\_fez} calibration snapshot
bundled with the installed package version. Hardware execution used
the \texttt{SamplerV2} primitive against the live \texttt{ibm\_fez}
backend via \texttt{QiskitRuntimeService}. Supporting analysis used
NumPy 2.0.2, SciPy 1.16.3, pandas 2.3.3, and matplotlib 3.10.0.
All code ran on Python 3.12.12.

\subsection{Device and Calibration}
\label{appendix:reproducibility:device}

Hardware experiments ran on \texttt{ibm\_fez}, a 156-qubit IBM Heron
r2 processor with heavy-hex connectivity and tunable couplers. The
device's native basis gates are $\{\texttt{cz},\,\texttt{id},\,
\texttt{rz},\,\texttt{sx},\,\texttt{x}\}$; all QuIC circuits are
transpiled into this gate set prior to execution. Hardware jobs for
the results reported in Sections~\ref{sec:hardware} and
~\ref{sec:hardware:extended} were executed between
27~March~2026 and 20~April~2026. A representative calibration
snapshot taken near the end of that window (backend properties
\texttt{last\_update\_date} 2026-04-20 18:10:47 UTC) gave median
$T_1 = 138.3~\mu$s (IQR $[106.5, 167.3]$), median $T_2 =
94.7~\mu$s (IQR $[48.1, 142.7]$), median single-qubit gate error
$2.68 \times 10^{-4}$, median two-qubit (CZ) gate error
$2.6 \times 10^{-3}$, and median readout error
$1.42 \times 10^{-2}$, measured across 156 qubits (312 single-qubit
gates, 352 CZ gates, 156 readout channels). Calibration drifts
between recalibrations, and all hardware claims in this paper should
be read as conditional on the calibration state during this
execution window; replication after substantial recalibration may
yield quantitatively different results even under an otherwise
identical execution protocol.

\subsection{Transpilation and Layout Selection}
\label{appendix:reproducibility:transpilation}

Transpilation targeted \texttt{ibm\_fez} using Qiskit's
\texttt{generate\_preset\_pass\_manager} at
\texttt{optimization\_level=3}, with explicit barriers inserted
between the encoding, entangling, and mixing layers to stabilize
compiler behavior; this barrier-stabilized configuration is the
execution protocol characterized in
Section~\ref{sec:hardware:transpilation}. For each CFI family, the
transpilation sweep used 20 edge orderings crossed with 10
\texttt{seed\_transpiler} values \texttt{range(10)} (i.e.,
$\{0,1,\dots,9\}$), producing 200 transpilations per family
(14{,}800 across 37 families in total). Hardware execution for path
and cycle families additionally used manual layouts excluding the
worst 30\% of two-qubit gate positions by device calibration at
submission time; this threshold was selected empirically as the
lowest exclusion level that consistently yielded feasible-depth
mappings across the family. All other families used the best-depth
automatic layout from the sweep.

\subsection{Shot Budget, Subsampling, and Statistics}
\label{appendix:reproducibility:statistics}

All hardware runs used $2^{15} = 32{,}768$ shots per circuit. Noise-model
and noiseless simulation runs used $2^{12} = 4{,}096$ shots unless
otherwise stated. For the $L_1$ $z$-score reported throughout the
practical sections, the null distribution is formed from 64
same-circuit subsamples of size $2^{12}$ drawn without replacement
from the full $2^{15}$-shot counts, and the signal distribution is
formed analogously from pairs of non-isomorphic circuits. The
$z$-score is
\[
z = \frac{\mu_{\mathrm{signal}} - \mu_{\mathrm{null}}}
         {\sigma_{\mathrm{pooled}}},
\]
with $\sigma_{\mathrm{pooled}}$ the pooled standard deviation across
null and signal samples. The operational separation criterion
throughout the practical sections is $z > 3$. Subsampling seeds were
fixed per experiment for within-run reproducibility.

\subsection{Dynamical Decoupling}
\label{appendix:reproducibility:dd}

Dynamical decoupling was applied via Qiskit's
\texttt{PadDynamicalDecoupling} transpilation pass using the XX
sequence ($X$--delay--$X$--delay), with \texttt{pulse\_alignment}
and \texttt{min\_delay} values inherited from the backend's
\texttt{InstructionDurations}. Pulses were inserted on idle qubits
only, and the pass was composed after layout selection and
scheduling. DD was used uniformly for all hardware runs reported in
Sections~\ref{sec:noisy} and~\ref{sec:hardware}.

\newpage

\section{Algorithm}
\label{appendix:algorithm}

Algorithm~\ref{alg:quic} gives the full QuIC embedding procedure as
a single reusable reference. Lines 3--5 implement the encoding
layer of Equation~(\ref{eq:encoding}); lines 6--13 the $r$-repetition
entangling--mixing structure of Equations~(\ref{eq:entangler})
and~(\ref{eq:mixer}); and lines 14--17 the sampling, sorting, and
head-truncation that produce the practical embedding. Under exact
arithmetic at $r = 1$ and the irrational-angle condition of
Section~\ref{sec:math:injectivity}, the returned $\tilde{p}$
(with $k = 2^n$) is the object shown to be a complete invariant
on isomorphism classes; all practical evaluations in this paper
correspond to finite-shot, fixed-$k$ executions of the same
procedure.

\begin{algorithm}[ht]
\caption{QuIC Graph Embedding}
\label{alg:quic}
\begin{algorithmic}[1]
\Require Graph $G = (V, E)$; angles $\theta_{\mathrm{enc}},
         \theta_{\mathrm{ent}}, \theta_{\mathrm{mix}}$;
         repetitions $r$; shots $N$; truncation $k$
\Ensure  Sorted truncated distribution $\tilde{p} \in \mathbb{R}^k$
\State $n \gets |V|$;\quad $d_{\max} \gets \max_{v \in V} \deg(v)$
\State Prepare $n$-qubit circuit in state $|0\rangle^{\otimes n}$
\For{$i = 1$ to $n$} \Comment{encoding}
    \State Apply $R_X\!\left(\theta_{\mathrm{enc}}\,\deg(v_i) / d_{\max}\right)$ to qubit $i$
\EndFor
\For{$\mathrm{round} = 1$ to $r$}
    \For{$(i, j) \in E$} \Comment{entangling}
        \State Apply $R_{ZZ}(\theta_{\mathrm{ent}})$ to qubits $(i, j)$
    \EndFor
    \For{$i = 1$ to $n$} \Comment{mixing}
        \State Apply $R_X(\theta_{\mathrm{mix}})$ to qubit $i$
    \EndFor
\EndFor
\State $\mathrm{counts} \gets$ sample circuit with $N$ shots
\State $p(x) \gets \mathrm{counts}(x) / N$ for all $x \in \{0,1\}^n$
\State $p_{\mathrm{sorted}} \gets \textsc{Sort}(p,\text{ descending})$
\State \Return $\tilde{p} \gets p_{\mathrm{sorted}}[1{:}k]$ \Comment{head truncation}
\end{algorithmic}
\end{algorithm}

\newpage

\section{Full Transpilation Depth Table}
\label{appendix:transpilation_depth}

Table~\ref{tab:transpilation_full} reports best- and worst-case
transpiled depths for all 37 CFI base graph families across the full
transpilation sweep of 14{,}800 circuits (20 edge orderings $\times$ 10
\texttt{seed\_transpiler} values $\in$ \texttt{range(10)} $\times$ 2
twist states), targeting \texttt{ibm\_fez} at
\texttt{optimization\_level=3} with layer-separating barriers. Depths
are transpiled serial depth in the native gate set
$\{\texttt{cz},\,\texttt{id},\,\texttt{rz},\,\texttt{sx},\,\texttt{x}\}$.
The ``range ratio'' column is the worst-case depth divided by the
best-case depth, taken over both twist states, and characterizes the
stability of each family under transpilation. Families are grouped by
structure and ordered by qubit count within each group.

\begin{table*}[ht]
    \centering
    \caption{Transpiled depth for all 37 CFI base graph families on
    \texttt{ibm\_fez}, 20 edge orderings $\times$ 10 seeds $\times$ 2
    twist states (14{,}800 circuits total). Best and worst depths per
    twist state; range ratio is worst/best across both states.
    Shaded rows exceed the 210--250 layer depth limit at best-case
    transpilation.}
    \label{tab:transpilation_full}
    \setlength{\tabcolsep}{6pt}
    \renewcommand{\arraystretch}{1.12}
    \small
    \begin{tabular}{llcccccc}
        \hline
        \textbf{Family} & \textbf{Base graph} & \textbf{Qubits}
        & \textbf{Untw best} & \textbf{Untw worst}
        & \textbf{Twist best} & \textbf{Twist worst}
        & \textbf{Range ratio} \\
        \hline
        \multirow{10}{*}{Path}
        & P$_3$        & 12 & 46   & 68   & 46   & 67   & 1.5$\times$  \\
        & P$_4$        & 18 & 48   & 94   & 49   & 89   & 2.0$\times$  \\
        & P$_5$        & 24 & 50   & 122  & 50   & 116  & 2.4$\times$  \\
        & P$_6$        & 30 & 51   & 149  & 51   & 137  & 2.9$\times$  \\
        & P$_7$        & 36 & 50   & 171  & 50   & 167  & 3.4$\times$  \\
        & P$_8$        & 42 & 56   & 195  & 50   & 184  & 3.9$\times$  \\
        & P$_9$        & 48 & 54   & 218  & 54   & 212  & 4.0$\times$  \\
        & P$_{10}$     & 54 & 54   & 238  & 54   & 229  & 4.4$\times$  \\
        & P$_{11}$     & 60 & 56   & 253  & 56   & 249  & 4.5$\times$  \\
        & P$_{12}$     & 66 & 53   & 280  & 53   & 273  & 5.3$\times$  \\
        \hline
        \multirow{8}{*}{Cycle}
        & C$_4$        & 24 & 50   & 99   & 48   & 96   & 2.1$\times$  \\
        & C$_5$        & 30 & 73   & 261  & 72   & 312  & 4.3$\times$  \\
        & C$_6$        & 36 & 80   & 377  & 50   & 149  & 7.5$\times$  \\
        & C$_7$        & 42 & 87   & 435  & 95   & 392  & 5.0$\times$  \\
        & C$_8$        & 48 & 55   & 187  & 55   & 198  & 3.6$\times$  \\
        & C$_9$        & 54 & 105  & 484  & 97   & 599  & 6.2$\times$  \\
        & C$_{10}$     & 60 & 95   & 800  & 59   & 248  & 13.6$\times$ \\
        & C$_{11}$     & 66 & 116  & 766  & 106  & 712  & 7.2$\times$  \\
        \hline
        \multirow{6}{*}{Pan}
        & Pan$_4$      & 25 & 138  & 358  & 142  & 331  & 2.6$\times$  \\
        & Pan$_5$      & 31 & 142  & 408  & 130  & 333  & 3.1$\times$  \\
        & Pan$_6$      & 37 & 146  & 380  & 150  & 394  & 2.7$\times$  \\
        & Pan$_7$      & 43 & 154  & 467  & 122  & 594  & 4.9$\times$  \\
        & Pan$_8$      & 49 & 149  & 527  & 144  & 643  & 4.5$\times$  \\
        & Pan$_9$      & 55 & 157  & 820  & 147  & 800  & 5.6$\times$  \\
        \hline
        \multirow{5}{*}{Complete}
        & K$_3$        & 18 & 63   & 160  & 56   & 162  & 2.9$\times$  \\
        & K$_4$-$e$    & 32 & 189  & 431  & 216  & 466  & 2.5$\times$  \\
        & K$_4$        & 40 & 341  & 717  & 340  & 705  & 2.1$\times$  \\
        & K$_5$-$e$    & 68 & 738  & 1579 & 724  & 1601 & 2.2$\times$  \\
        & K$_5$        & 80 & 952  & 1829 & 979  & 1884 & 2.0$\times$  \\
        \hline
        \multirow{4}{*}{Star}
        & Star$_4$     & 19 & 120  & 272  & 124  & 250  & 2.3$\times$  \\
        & Star$_5$     & 28 & 360  & 667  & 365  & 603  & 1.9$\times$  \\
        & Star$_6$     & 41 & 937  & 1467 & 922  & 1426 & 1.6$\times$  \\
        & Star$_7$     & 62 & 1873 & 3199 & 1772 & 3215 & 1.8$\times$  \\
        \hline
        \multirow{4}{*}{Bipartite}
        & K$_{2,3}$    & 38 & 187  & 455  & 211  & 489  & 2.6$\times$  \\
        & K$_{2,4}$    & 56 & 498  & 1019 & 497  & 964  & 2.0$\times$  \\
        & K$_{3,3}$    & 60 & 382  & 769  & 426  & 768  & 2.0$\times$  \\
        & K$_{3,4}$    & 88 & 819  & 1445 & 837  & 1596 & 2.0$\times$  \\
        \hline
    \end{tabular}
\end{table*}

Several structural patterns follow directly from the table. Path
families stay below the 210--250 layer depth limit throughout the
full 12--66 qubit range at best-case transpilation, reaching a best
depth of 53 at 66 qubits. Cycle families are likewise feasible at all
tested sizes under best-case transpilation, though their worst-case
depths exhibit the largest range ratios in the sweep
(CFI($C_{10}$) at 13.6$\times$, CFI($C_{11}$) at 7.2$\times$, and
CFI($C_6$) at 7.5$\times$). Pan families cross the depth limit in the
worst case but remain feasible at best-case transpilation through
Pan$_9$. Complete, star, and denser bipartite families exceed the
depth limit at their smallest instances and scale rapidly: CFI($K_4$)
at 40 qubits has a best depth of 340, already outside the feasible
regime, and CFI(Star$_7$) at 62 qubits has best depths near 1{,}800
layers even before considering the worst-case tail. These per-family
patterns are the source of the feasibility-by-density finding
reported in Section~\ref{sec:hardware:transpilation}.

\newpage

\section{Mathematical Proofs}
\label{appendix:proofs}
All claims in this section concern the exact noiseless output
distribution induced by the ideal circuit and do not involve sampling,
truncation, transpilation, or hardware noise. Throughout this section,
we consider simple undirected unlabeled graphs.

\subsection{Permutation Invariance}\label{appendix:proofs:invariance}
For a graph $G$, let $\tilde{p}(G)$ denote the sorted exact output
distribution defined in Section~\ref{sec:construction}.
\begin{theorem}
Let $G$ and $H$ be isomorphic simple undirected graphs. Then
\[
\tilde{p}(G)=\tilde{p}(H).
\]
\end{theorem}

\begin{proof}
Let $G=(V,E)$ and let $H=\pi(G)$ be obtained from $G$ by a relabeling of
the vertices under a permutation $\pi$ of $\{1,\dots,k\}$. Under the
QuIC construction, relabeling the vertices induces the same permutation
of the qubit indices.

The encoding layer depends only on the multiset of vertex degrees, so a
vertex relabeling permutes the single-qubit $R_X$ gates accordingly. The
entangling layer is defined by the edge set, and a relabeling maps each
edge $(i,j)\in E(G)$ to the corresponding edge $(\pi(i),\pi(j))\in E(H)$.
Since the $R_{ZZ}$ gates are order-invariant, this relabeling induces
only a permutation of qubit positions, not a change in the underlying
operator family. The mixing layer is uniform across all qubits and is
therefore unchanged by relabeling.

It follows that $U(H)$ is obtained from $U(G)$ by conjugation with the
qubit-permutation operator induced by $\pi$. Measuring in the
computational basis therefore yields the same probability values, up to a
permutation of the bitstring labels. Sorting removes this residual
dependence on basis ordering, so the sorted output distributions are
identical:
\[
\tilde{p}(G)=\tilde{p}(H).
\]
\end{proof}

\subsection{Injectivity of the Circuit}\label{appendix:proofs:injectivity}
We first record a reconstruction lemma showing that the full cut
function determines the graph on a fixed labeled vertex set.

\begin{lemma}[Cut reconstruction]
Let $G=([n],E)$ and $H=([n],E')$ be simple undirected graphs on the same
labeled vertex set. If
\[
\mathrm{cut}_G(s)=\mathrm{cut}_H(s)
\qquad\text{for all } s\in\{0,1\}^n,
\]
then $E=E'$.
\end{lemma}

\begin{proof}
For $s\in\{0,1\}^n$, the cut function may be written as
\[
\mathrm{cut}_G(s)
=
\sum_{(u,v)\in E}\mathbf{1}[s_u\neq s_v]
=
\sum_{(u,v)\in E}(s_u+s_v-2s_us_v).
\]
This is a multilinear polynomial in the Boolean variables
$s_1,\dots,s_n$. The coefficient of the monomial $s_u s_v$ is $-2$ if
$(u,v)\in E$ and $0$ otherwise. Since multilinear polynomials are
uniquely determined by their values on $\{0,1\}^n$, equality of cut
functions implies equality of all coefficients, hence $E=E'$.
\end{proof}

The next theorem establishes injectivity for the single-repetition
statevector.

\begin{theorem}[One-repetition statevector injectivity]
Let $G=([n],E_G)$ and $H=([n],E_H)$ be simple undirected graphs on the
same labeled vertex set. Let
\[
U_K = Z\,Y_K\,X_K,
\]
where $X_K$ is the degree-encoding layer for graph $K$, $Y_K$ is the
entangling layer for graph $K$, and $Z$ is the common mixer. Assume
\[
\theta_{\mathrm{enc}},\theta_{\mathrm{ent}},\theta_{\mathrm{mix}}
\notin \pi\mathbb{Q},
\]
and in particular that $\theta_{\mathrm{enc}}<\pi$ so that the encoding
angles lie in $[0,\pi/2)$. If
\[
U_G\lvert 0\rangle^{\otimes n}
=
U_H\lvert 0\rangle^{\otimes n},
\]
then $G=H$ as labeled graphs.
\end{theorem}

\begin{proof}
Let
\[
\lvert \phi_K\rangle := X_K\lvert 0\rangle^{\otimes n}.
\]
Since $Z$ is a common unitary, equality of final states implies
\[
Y_G\lvert \phi_G\rangle = Y_H\lvert \phi_H\rangle.
\]
For each $K\in\{G,H\}$, write
\[
Y_K\lvert s\rangle = \lambda_K(s)\lvert s\rangle,
\qquad s\in\{0,1\}^n,
\]
where $|\lambda_K(s)|=1$. Expanding in the computational basis gives
\[
\lambda_G(s)\,\langle s\mid \phi_G\rangle
=
\lambda_H(s)\,\langle s\mid \phi_H\rangle
\qquad\forall s\in\{0,1\}^n.
\]
Taking moduli yields
\[
|\langle s\mid \phi_G\rangle|
=
|\langle s\mid \phi_H\rangle|
\qquad\forall s\in\{0,1\}^n.
\]

Now $\lvert\phi_K\rangle$ is a product state. Writing
\[
\alpha_i^K
=
\frac{\theta_{\mathrm{enc}}}{2}\,
\frac{\deg_K(i)}{\max_j \deg_K(j)},
\]
with the convention $\alpha_i^K=0$ when $\deg_K(i)=0$, we have
\[
\langle s\mid \phi_K\rangle
=
\prod_{i:s_i=0}\cos(\alpha_i^K)\,
\prod_{i:s_i=1}\bigl(-i\sin(\alpha_i^K)\bigr).
\]
Since $\alpha_i^K\in[0,\pi/2)$, both $\cos(\alpha_i^K)$ and
$\sin(\alpha_i^K)$ are nonnegative, and therefore
\[
|\langle s\mid \phi_K\rangle|
=
\prod_{i:s_i=0}\cos(\alpha_i^K)\,
\prod_{i:s_i=1}\sin(\alpha_i^K).
\]

Let
\[
c_i^K := \cos(\alpha_i^K),
\qquad
r_i^K :=
\begin{cases}
\tan(\alpha_i^K), & c_i^K>0,\\
0, & c_i^K=0,
\end{cases}
\]
so that
\[
|\langle s\mid \phi_K\rangle|
=
\left(\prod_{i=1}^n c_i^K\right)
\prod_{i:s_i=1} r_i^K.
\]
Equality of these quantities for every bitstring $s$ implies first, by
taking $s=0^n$, that
\[
\prod_{i=1}^n c_i^G = \prod_{i=1}^n c_i^H.
\]
Then taking $s=e_i$ (the bitstring of Hamming weight one at position
$i$) gives
\[
r_i^G = r_i^H
\qquad\forall i.
\]
Hence
\[
\tan(\alpha_i^G)=\tan(\alpha_i^H)
\qquad\forall i.
\]
Because $\tan$ is injective on $[0,\pi/2)$, it follows that
\[
\alpha_i^G=\alpha_i^H
\qquad\forall i,
\]
and therefore
\[
\deg_G(i)=\deg_H(i)
\qquad\forall i.
\]
Thus the encoding layers coincide:
\[
X_G = X_H,
\qquad
\lvert\phi_G\rangle = \lvert\phi_H\rangle.
\]

Returning to the componentwise equality, we now obtain
\[
\lambda_G(s)=\lambda_H(s)
\qquad\forall s\in\{0,1\}^n
\]
on every basis state for which the common encoder amplitude is nonzero.
Using the standard diagonal form of the entangler,
\[
\lambda_K(s)
=
\exp\!\left(i\theta_{\mathrm{ent}}\,
\mathrm{cut}_K(s)\right)\,
\exp\!\left(-i\frac{\theta_{\mathrm{ent}}}{2}\,|E_K|\right),
\]
so
\[
\exp\!\left(
i\theta_{\mathrm{ent}}
\bigl(\mathrm{cut}_G(s)-\mathrm{cut}_H(s)\bigr)
\right)
=
\exp\!\left(
i\frac{\theta_{\mathrm{ent}}}{2}
\bigl(|E_G|-|E_H|\bigr)
\right),
\]
where the right-hand side is independent of $s$. Thus for any
$s,t\in\{0,1\}^n$,
\[
\exp\!\left(
i\theta_{\mathrm{ent}}
\bigl[
\bigl(\mathrm{cut}_G(s)-\mathrm{cut}_H(s)\bigr)
-
\bigl(\mathrm{cut}_G(t)-\mathrm{cut}_H(t)\bigr)
\bigr]
\right)
=1.
\]
The quantity in brackets is an integer, and
$\theta_{\mathrm{ent}}/2\pi\notin\mathbb{Q}$ by assumption. Therefore
the only possibility is
\[
\bigl(\mathrm{cut}_G(s)-\mathrm{cut}_H(s)\bigr)
-
\bigl(\mathrm{cut}_G(t)-\mathrm{cut}_H(t)\bigr)
=0
\qquad\forall s,t.
\]
Hence
\[
\mathrm{cut}_G(s)-\mathrm{cut}_H(s)
\]
is constant as a function of $s$. Evaluating at $s=0^n$, where both cut
values are zero, shows that this constant is zero. Therefore
\[
\mathrm{cut}_G(s)=\mathrm{cut}_H(s)
\qquad\forall s\in\{0,1\}^n.
\]
By the cut-reconstruction lemma, $G=H$.
\end{proof}

\begin{corollary}
For one repetition, the ideal statevector map
\[
G \mapsto Z\,Y_G\,X_G\lvert 0\rangle^{\otimes n}
\]
is injective on labeled simple undirected graphs.
\end{corollary}

\begin{corollary}[Complete invariant on isomorphism classes]
For one repetition, the sorted ideal output distribution $\tilde{p}(G)$ is a complete invariant on isomorphism classes of simple undirected graphs: $\tilde{p}(G) = \tilde{p}(H) \iff G\cong H$.
\end{corollary}

\begin{proof}
The forward direction is the permutation invariance theorem of Appendix \ref{appendix:proofs:invariance}. For the reverse, suppose $G\not\cong H$. Pick any labelings of $G$ and $H$ on $[n]$. By the labeled injectivity theorem, their ideal statevectors differ, hence so do their unsorted Born distributions. Since each isomorphism class produces a single sorted distribution under permutation invariance, and the two classes are distinct, their sorted distributions differ.
\end{proof}

\begin{remark}
The theorem above concerns equality of ideal statevectors, not equality of measured output distributions. Distribution-level distinguishability and finite-shot estimation are treated separately below.
\end{remark}

\begin{conjecture}[Multi-repetition injectivity on a common-encoder slice]
Let $G$ and $H$ be simple undirected graphs on the same labeled vertex
set, and suppose their encoding layers coincide:
\[
X_G = X_H =: X.
\]
For any repetition count $r\geq 1$, define
\[
U_K = (Z\,Y_K)^r X.
\]
Under the same irrational-angle assumptions, we conjecture that
\[
U_G\lvert 0\rangle^{\otimes n}
=
U_H\lvert 0\rangle^{\otimes n}
\quad\Longrightarrow\quad
G=H.
\]
\end{conjecture}

\begin{remark}[Supporting intuition]
Write
\[
\lvert s_0^K\rangle = X\lvert 0\rangle^{\otimes n},
\qquad
\lvert t_k^K\rangle = Y_K\lvert s_{k-1}^K\rangle,
\qquad
\lvert s_k^K\rangle = Z\lvert t_k^K\rangle,
\]
for $k=1,\dots,r$. If the final states coincide,
\[
\lvert s_r^G\rangle = \lvert s_r^H\rangle,
\]
then invertibility of the common mixer gives
\[
\lvert t_r^G\rangle = \lvert t_r^H\rangle.
\]
Since each entangler is diagonal in the computational basis, this
induces a diagonal rephasing relation between the preceding states:
\[
\langle s \mid s_{r-1}^G\rangle
=
\exp\!\left(
i\theta_{\mathrm{ent}}
\bigl(\mathrm{cut}_H(s)-\mathrm{cut}_G(s)\bigr)
\right)
\langle s \mid s_{r-1}^H\rangle.
\]
Repeating this backward peeling through successive mixer layers yields a
recurrence of graph-dependent diagonal rephasings. Because the mixer
does not preserve nontrivial diagonal operators in the computational
basis, and because the irrational-angle condition prevents periodic
phase aliasing, the only consistent solution appears to be the trivial
one in which all rephasings are identity and hence all cut differences
vanish. This is strongly supported by the numerical evidence, but a full
proof for arbitrary $r$ remains open.
\end{remark}

\end{document}